\documentclass[aps,twocolumn,reprint,letterpaper,showkeys]{revtex4-2}

\usepackage{graphicx}
\usepackage{multirow}

\newcommand{\fone}{120}

\newcommand{\ffour}{50}

\begin{document}

\title{Ice friction at the nanoscale}


\author{Lukasz Baran}
\affiliation{Department of Theoretical Chemistry, Institute of Chemical Sciences, Faculty
	of Chemistry, Maria-Curie-Sklodowska University in Lublin, Pl. M
	Curie-Sklodowskiej 3, 20-031 Lublin, Poland.}
\author{Pablo Llombart} 
\affiliation{Departamento de F\'{\i}sica Te{\'o}rica de la Materia Condensada, Instituto Nicol{\'a}s Cabrera, Universidad Aut{\'o}noma de Madrid, 28049 Madrid, Spain}
\author{Wojciech R{\.z}ysko} 
\affiliation{Department of Theoretical Chemistry, Institute of Chemical Sciences, Faculty
	of Chemistry, Maria-Curie-Sklodowska University in Lublin, Pl. M
	Curie-Sklodowskiej 3, 20-031 Lublin, Poland.}
\author{Luis G. MacDowell}
\affiliation{Departamento de Qu\'{\i}mica-F\'{\i}sica, Facultad de Ciencias Qu\'{\i}micas, Universidad Complutense de Madrid, 28040 Madrid, Spain.}
\email{Corresponding author: lgmac@quim.ucm.es}


\begin{abstract}
The origin of ice slipperiness has been a matter of great controversy for
more than a century, but an atomistic understanding of ice friction
is still lacking. Here, we perform   computer simulations of an atomically
smooth substrate sliding on ice. In a large temperature range between 230 and
266~K, hydrophobic sliders exhibit  a premelting layer
similar to that found at the ice/air interface.  On the contrary,
hydrophilic sliders show larger premelting and a strong increase of the first
adsorption layer. The non-equilibrium simulations show that premelting films of barely one nanometer thickness are sufficient  to provide a lubricating quasi-liquid layer with rheological
properties similar to bulk undercooled water. Upon shearing,  the films
display  a pattern consistent with lubricating Couette flow,  but the
boundary conditions at the wall vary strongly with substrate’s interactions.
Hydrophobic walls exhibit large slip, while hydrophilic walls obey stick
boundary
   conditions with small negative slip.  By compressing ice above atmospheric
pressure, the lubricating
   layer grows continuously, and the rheological properties approach bulk--like
   behavior.  Below 260 K, the equilibrium premelting films decrease
significantly. However,
a very large slip persists on the hydrophobic walls, while the increased
friction on hydrophilic walls is sufficient to melt ice and create a lubrication
layer in a few nanoseconds.  Our results show the atomic scale frictional behavior of ice 
is a combination of spontaneous  premelting, pressure melting and frictional
   heating.
\end{abstract}


\maketitle

The slipperiness of ice has been exploited since ancient times as a
means of transportation in cold regions \cite{li13b}. 

But despite  many advances on tribology
\cite{robbins00,persson00,vanossi13}
a first principles understanding on this very familiar property is still lacking
\cite{rosenberg05}.

A hypothesis dating back to the 19th century is that a
self-lubricating water film on the ice surface is formed due to pressure 
melting \cite{joly86,reynolds99}.
Spontaneous equilibrium
premelting \cite{jellinek67}, and frictional heating \cite{bowden53}, 
have also been invoked to explain ice
friction. However, other authors disregard the significance of water lubrication
altogether \cite{budnevich94,tusima11,weber18,liefferink21}, while 
recent experiments support boundary  or elastohydrodynamic models of
friction \cite{weber18,canale19,liefferink21,lever21b}.
Experimental confirmation of interfacial premelting films in the order of the
nanometer does not resolve the
controversy \cite{dash99,beaglehole94,wettlaufer99b,engemann04,liljeblad17,li21}, as  it is arguable 
whether macroscopic hydrodynamics assumed in most
theories \cite{oksanen82,colbeck88,lozowski13} is obeyed at such small
length-scales \cite{majumder05}.
In fact, computer simulations of flow under confinement reveal 
consistent violation of the stick boundary condition and the significance of 
water slip \cite{falk10,hansen11,kannam13,dilecce20,herrero20}, while  studies of water sliding on ice 
and grain boundary friction suggest negative slip 
instead \cite{louden18,ribeiro21}. 

Here we report Molecular Dynamics simulations of ice sliding 
past an atomically
smooth substrate. Our results show that an interfacial premelting
film  formed spontaneously  upon compression  or frictional heating, exhibits hydrodynamic
properties similar to bulk undercooled water.  This illustrates that a
hydrodynamic theory of Couette flow supplemented with slip  boundary conditions
can explain the friction of ice at smooth contacts.

\section*{Results}

\begin{figure}[t]
   \centering
   \includegraphics[width=0.48\textwidth]{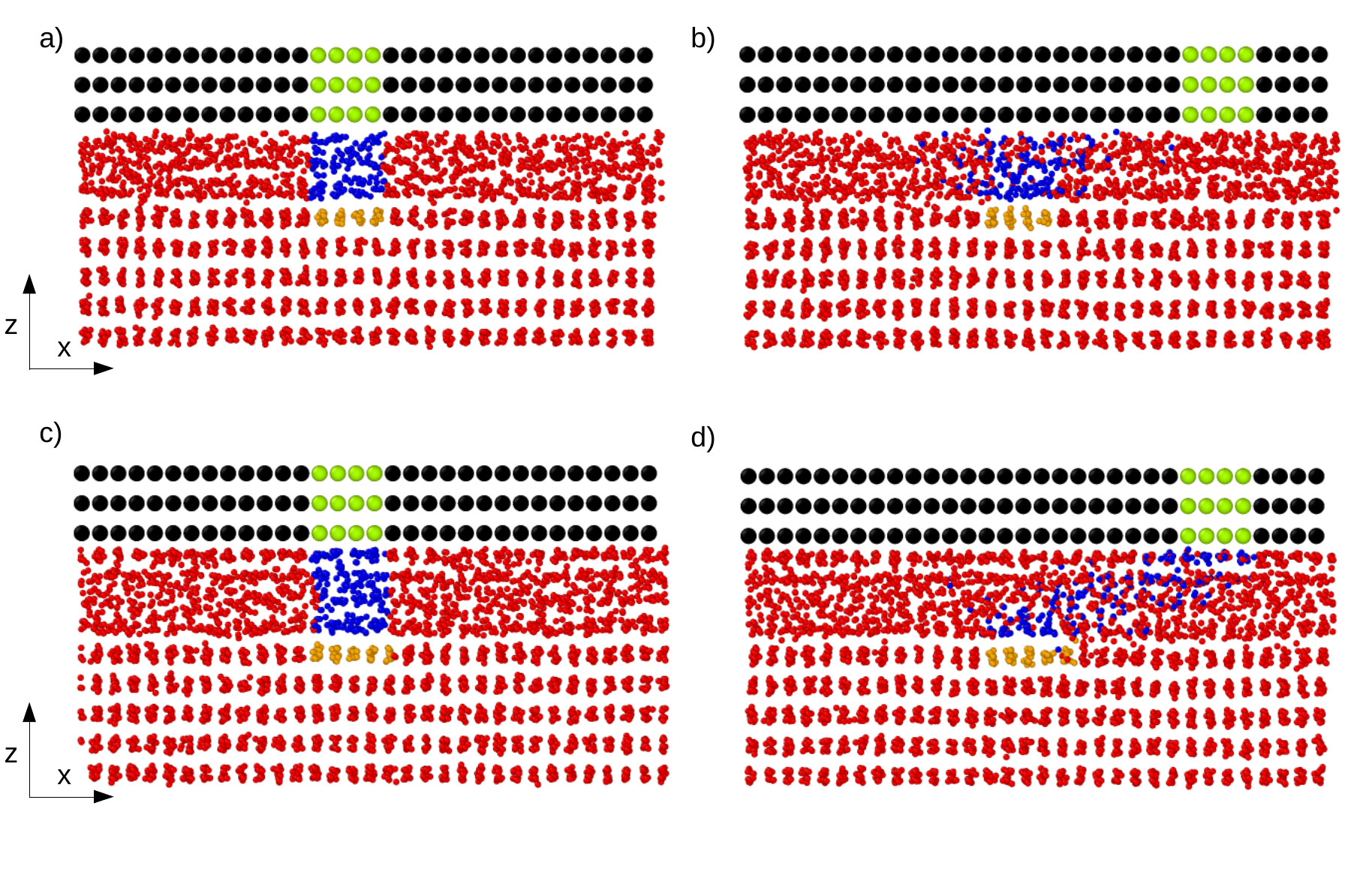}
   \caption{
	Sliding on ice with atomic resolution. 
	An imperfectly terminated ice slab gently compressed to $p=1$~atm by
	an inert solid substrate at T=262~K spontaneously develops a premelting film of 
	thickness ca. $h=1$~nm for both hydrophobic \textbf{(a)} and hydrophilic walls \textbf{(c)}. 
	The flow pattern after sliding over the
	equilibrated film  with sliding velocity $U=5$~m$/$s during 0.5~ns
	is illustrated by liquid-like molecules tagged in blue and wall atoms tagged in green color at $t=0$.
	\textbf{(b)} For the hydrophobic substrate, the slider slips
	past the premelting film, and the liquid-like blue tagged molecules
	have hardly moved beyond their original position at $t=0$
	(see also Movie S1).
	\textbf{(d)} For the hydrophilic substrate, an
	adsorption layer sticks to the wall and the premelting film is dragged by
	the slider with a pattern that resembles Couette flow (see also Supplementary Movie 2). Solid like orange colored
	molecules serve to illustrate stick boundary conditions
	at the ice/water interface and the exchange of solid and liquid molecules.
   }
   \label{Fsnapshotsp1}
\end{figure}

In our study, we simulate explicitly ice sliding past an
atomically smooth substrate under pressure (Fig.~\ref{Fsnapshotsp1}). 
The ice sample consists of a large orthorombic slab of water
molecules (30 bilayers thick)
modeled with the TIP4P/Ice force field and oriented in the direction of the
basal surface \cite{abascal05}.
The slider is modeled as a 
rigid face--centered cubic arrangement of atoms directed 
along the (111) plane, with lattice parameters selected to make a perfect 
match with the ice surface. 

The wall atoms interact with water oxygens
via a Lennard-Jones potential. This allows us to tune the hydrophobicity 
of the substrate merely by changing the strength of wall-oxygen interactions.
The quality of the substrate is monitored by measuring the contact angle of
water droplets, $\theta$,
which is varied in a range spaning both hydrophobic ($\theta=120^\circ$)
and  hydrophilic ($\theta=50^\circ$) walls (SI Appendix Text and Methods).

\begin{figure*}[t]
   \centering
   \includegraphics[width=0.95\textwidth]{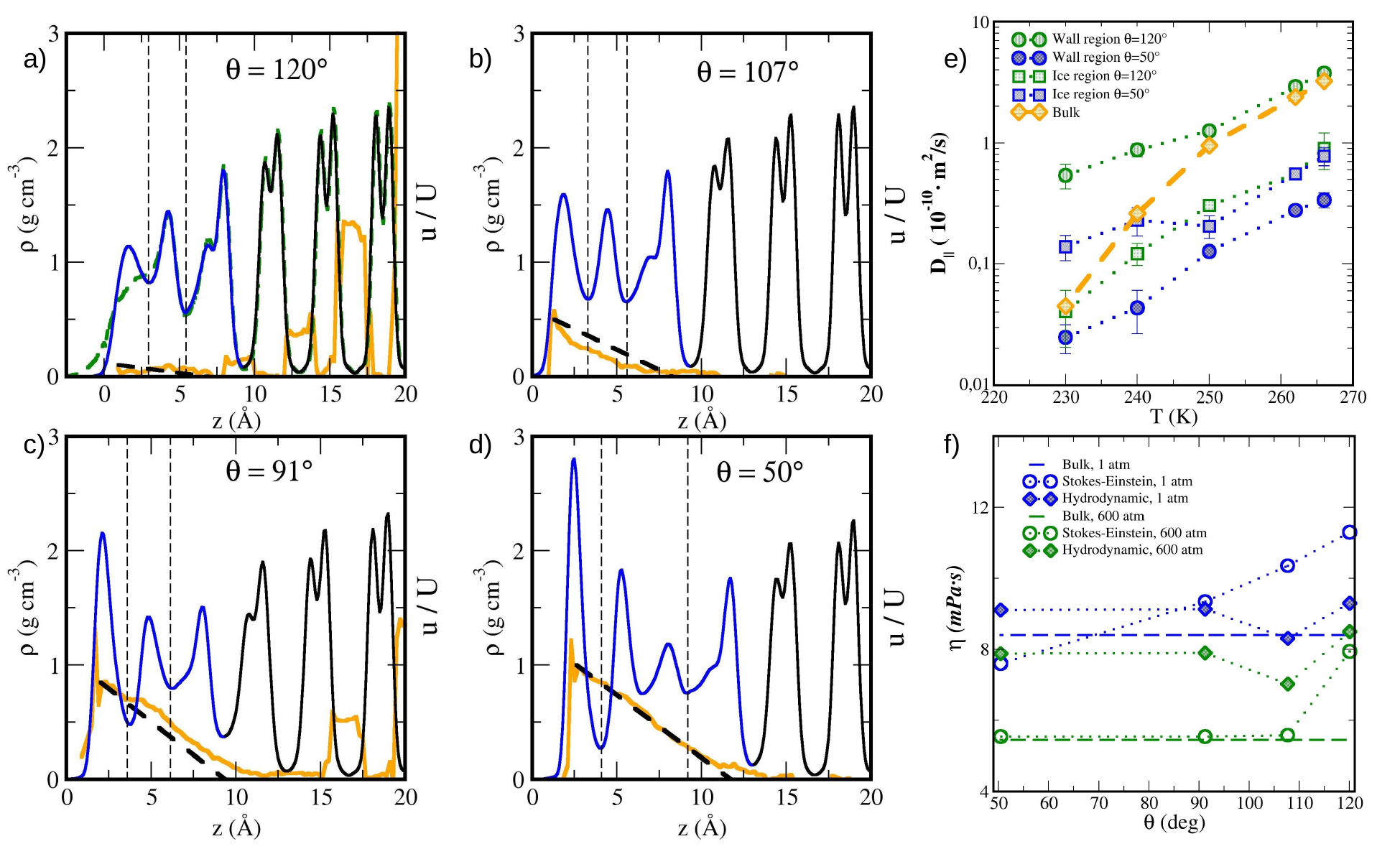}
   \caption{
	Structure and dynamics of premelted films
	during sliding. Panels ({\bf a,b,c,d}) show equilibrium density profiles
	and velocity profiles for a a sliding experiment at $p=1$~atm, $T=262$~K
	and $U=5$~m$/s$ during 10~ns.  
	The total density profile (left axis) is shown as a continuous line,
	with blue colour for the region where liquid-like water is the majority
	phase and black colour where ice is the majority phase. 
	The premelting film is divided into a wall-adsorbed layer, an ice-adsorbed
	layer and a central quasi-bulk region, as illustrated by
	vertical dashed lines.
	Panel (a) also shows the total density of a premelting film for
	ice in contact with vacuum (green dashed line).  The thick
	orange lines display the velocity profile in units of
	the sliding velocity (right axis, with tick marks as displayed in the left axis).  
      The dashed black line is the hydrodynamic flow
	profile predicted from the model of Eq.~(\ref{eq:model}).  Panels correspond
	to different wall interactions  (a) Hydrophobic wall, with 
	$\theta=120^\circ$ (b) $\theta=107^\circ$ (c)
	$\theta=91^\circ$ (d) Hydrophilic
	wall with $\theta=50^\circ$.
	Panel ({\bf e}): 
	Diffusivity at the wall and ice adsorption layers as a function of temperature.
	The orange diamonds 
	stand for the bulk diffusion coefficients at $p=1$~atm. 
	The remaining symbols correspond to parallel diffusion coefficients of
	the wall-adsorption layer (circles) and the ice-Adsorption layer (squares);
	Green symbols stand for the hydrophobic wall with $\theta=120^\circ$ and
   blue symbols for the hydrophilic wall with $\theta=50^\circ$.
	Panel ({\bf f}): Quasi-bulk like viscosity of the premelting film.
	The thick dashed line displays the shear viscosity of bulk undercooled
	water as determined from Green-Kubo calculations. The circles display
	viscosities as determined from the Stokes-Einstein relation using the
	parallel diffusion coefficient calculated in the central quasi-bulk like
	region of the premelting films. The diamonds are hydrodynamic viscosities
	as determined from the film thickness and shear stress of the simulations.
	Results are shown for $T=262$~K, with  $p=1$~atm (blue) and $p=600$~atm (green).
   }
   \label{Fig2}
\end{figure*}

Under skating conditions, the slider does not step on a perfectly terminated ice
surface. Instead, the ice surface has been previously exposed to air, and
exhibits a significant premelting layer
\cite{kling18,louden18,llombart20b,slater19}.
To mimic the contact of the slider with ice, we merely prepare the ice surface with  
a half terminated bilayer and place it at a small distance from the substrate.
The wall is then allowed to gently compress the
slab to the desired pressure (see Methods Section).  At a temperature $T=262$~K, somewhat lower but
close to that of skating rinks, we find that a 
premelting film of the order of a nanometer thick evolves spontaneously and equilibrates in the scale of 
decades of nanoseconds for all substrates and pressures studied.
The presence of premelting is obvious in the snapshots of Fig.~\ref{Fsnapshotsp1} as a layer of disordered
water molecules between the ordered bulk ice and the
substrate. In the density
profiles of Fig.~\ref{Fig2}-a,b,c,d, the signature of premelting is the emergence of density peaks
that have lost the bilayer structure typical of bulk ice that is apparent
within the bulk region. This is confirmed by use of the CHILL+ order parameter \cite{nguyen15}, which allows to resolve solid-like from liquid-like water molecules
	(SI Appendix Methods).

After equilibration, we model sliding by moving the top and bottom
sliders  with equal sliding speed $U=5$~m$/$s and opposite direction.  
Although the properties of premelting layers of ice exposed to vacuum are often
invoked as a proxy to explain ice friction \cite{rosenberg05,louden18,weber18},
inspection of simulation snapshots show  a dramatic dependence of the sliding
dynamics on the substrate interactions. Here we describe results obtained
at ambient pressure $p=1$~atm and $T=262$~K   (Fig.~\ref{Fsnapshotsp1} and movies S1 and S2), but similar results are found in all the temperature	range from 230 to 266~K (c.f. SI Appendix, Fig. S1 and S2).

For the hydrophobic substrate, $\theta=120^\circ$, the  slider slips past the
premelting film, and generates an extremely small velocity field.
Molecules tagged in blue in Fig.~\ref{Fsnapshotsp1}-(a) at $t=0$ 
have diffused almost
equally in both directions after a sliding time of 0.5~ns (Fig.~\ref{Fsnapshotsp1}-(b)). i.e.: as observed in the flow of water inside carbon
nanotubes \cite{falk10,kannam13,herrero19,herrero20}, friction is extremely 
small, and the premelting film is hardly susceptible to the motion of the 
slider. 
On the contrary, for the hydrophilic substrate, with
$\theta=50^\circ$,
an adsorbed layer of water molecules next to the 
substrate at $t=0$ (Fig.~\ref{Fsnapshotsp1}-(c)) sticks to the wall and is
displaced  by the same amount as the slider after 0.5~ns  (Fig.~\ref{Fsnapshotsp1}-(d)).  
The remaining blue tagged molecules in the premelting film are loosely dragged
by the slider and display clear hints of  Couette flow. 

The large difference in the frictional behavior can be anticipated
from the plot of equilibrium density profiles \cite{nikifioridis21}. Here we
describe 	results obtained at T=262~K (Fig.~\ref{Fig2}-(a,b,c,d)), but the
same trend is observed in all the temperature range studied (SI Appendix Fig, S3
and S4). For the hydrophobic substrate ($\theta=120^\circ$) the structure of
the  premelting film is very similar to that found when the ice
surface is exposed to vacuum (green dashed line in
Fig.~\ref{Fig2}-(a)). The density profiles  differ significantly
only by the presence of a small density peak that appears in the
confined film close to the wall.  However, increasing the strength
of the wall interactions results in an increase of the film thickness
and the appearance of a 
strongly layered liquid film.
Particularly, we see a large enhancement of the first adsorption peak,
with a density that can increase as much as a factor of three compared to
that observed in the hydrophobic wall with $\theta=120^\circ$ (Fig.~\ref{Fig2}-(d)). By visual
inspection we confirm that the water molecules pertaining to the first
adsorption peak exhibit strong intra-layer hydrogen bonding, with proliferation
of flattened hexagonal rings as observed in adsorbed thin films on metals \cite{carrasco12,shimizu18}, and  undercooled water under
confinement \cite{kastelowitz10,zhu19}.

We expect the hydrogen bond network on the first adsorption peak of hydrophilic
substrates to have a significant impact on the mobility of water
molecules \cite{louden18,kling18,nikifioridis21}.  
To show this, 
we divide the premelting film into regions that allow us to single out the wall adsorption layer from the ice adsorption layer, as illustrated with vertical dashed lines in  Fig.~\ref{Fig2}-(a,b,c,d)).
For each of these two regions, we estimate an effective parallel self-diffusion coefficient, $D_{\parallel}$
by measuring the tangential mean squared displacement 
of the water like molecules in
that region (Fig.~\ref{Fig2}-(e)). Our results 
confirm a dramatic impact of the wall-water
interactions on the mobility of the wall-adsorption layer. For the hydrophobic
substrate, $\theta=120^\circ$, the parallel diffusion coefficient is somewhat
larger than
that of bulk undercooled water  for most temperatures studied, as observed in
premelting films exposed to vacuum \cite{weber18,louden18,kling18}, and becomes an order of magnitude larger on approaching 230~K.
However, for the hydrophilic substrate, $\theta=50^\circ$,  the diffusion coefficient of the wall adsorption layer remains one order of magnitude smaller than that of the hydrophobic substrate all the way from 266 to 230~K.
On the other hand, for the ice-adsorption layer the diffusion coefficient remains small and almost independent of $\theta$, 
implying that the details of the slider do not impact the friction of
premelted water at the ice interface.

For temperatures above 260~K, the premelting layer develops a well defined quasi-bulk region between the adsorption layers, as observed in Fig.2 and SI Appendix Figs. S3 and S4.  Our results show that the parallel diffusion coefficient in this central region is somewhat smaller, 
but of the same
order of magnitude as the diffusion coefficient of bulk water, even for the
films studied here, which are barely 1~nm thick \cite{falk10} (c.f. SI Appendix Fig. S5).  This is in
agreement with measurements of mobility in confined 
water \cite{dash95,falk10,xu16,li21}, and suggests that
large effective viscosities measured in mechanical tests  \cite{canale19}
might not be 
related to the actual hydrodynamics of the premelting film,  
as noticed 
in Ref.~\cite{li21}.
	This is not in conflict with the observation of anomalous diffusion at
	grain boundaries, which is related to the  attachment/detachment of water
	molecules into the ordered ice lattice due to motion in the \emph{perpendicular} direction  \cite{moreira18}. 
We find that in the time scale of about 2-8~ns in which
premelted water molecules remain within the central quasi-bulk region of the 
premelting
film, the mobility in the  \emph{parallel direction} remains close to 
bulk like. 
This  suggests that the central region within the premelting
film will 
exhibit shearing similar to that observed in bulk undercooled water. As a hint,
we notice that the viscosity predicted from the Stokes-Einstein relation
$\eta_{SE}=k_BT/6\pi D_{\parallel} a$, with $a=0.155$~nm describing the
molecular radius of a water molecule provides
an order of magnitude approximation to
the viscosity of undercooled water calculated independently
at $T=262$~K and $p=1$~atm from bulk simulations (Fig.~\ref{Fig2}-(f)). 

\begin{figure}[t]
   \centering
   \includegraphics[width=0.4\textwidth,trim=6cm 2cm 6cm 10cm]{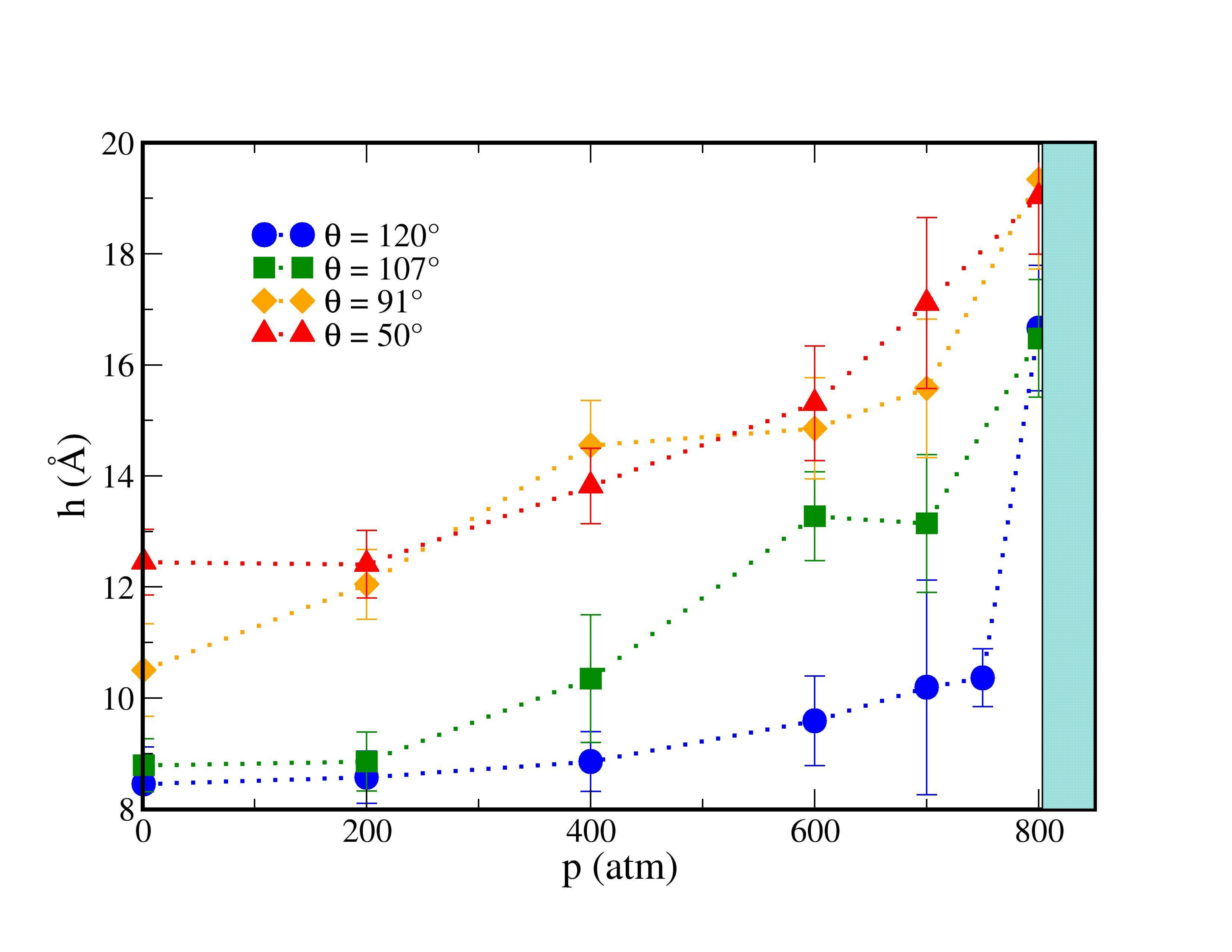}
   \caption{
	Increasing equilibrium premelting thickness by compression. 
	Results display the equilibrium interfacial premelting thickness as a
	function of pressure for different substrates at $T=262$~K. Blue circles:
	$\theta=120^\circ$; Green squares: $\theta=107^\circ$; organge triangles:
	$\theta=91^\circ$; red triangles: $\theta=50^\circ$. The hashed region
	displays the estimated melting pressure for the model.
   } 
   \label{Fthickness}
\end{figure}

In practice, sliding occurs at contact with surface asperities, and the
pressure on the contact zone can well reach several hundred
atmospheres \cite{bowden53,colbeck88,lozowski13,weber18,liefferink21,lever21b}.
However, increasing pressure drives ice closer to the melting line. Therefore,
the thickness of the premelting film is expected to increase.
We confirm this by compressing our confined ice slabs, and estimating
the equilibrium film thickness as $h_{eq}=\Gamma_w/\rho_w$, with $\Gamma_w$,
the number of liquid--like molecules per unit surface, and $\rho_w$ the bulk
liquid density. The results
of Fig.~\ref{Fthickness} show that, independent of the substrate quality, the
interfacially premelted films increase their thickness under compression,
showing that pressure melting and interfacial premelting are inextricably
entangled \cite{dash99,wettlaufer99b}. 

As a result of this surface-pressure melting, the diffusion coefficient
and the corresponding Stokes-Einstein viscosity of the films approach
bulk-like conditions as illustrated in SI Appendix Fig. S5 
and Fig.~\ref{Fig2}-(f) for interfacially premelted films compressed
at a pressure of $p=600$~atm.

In order to check how bulk-like is the sliding hydrodynamics of a quasi-liquid layer
barely one nanometer thick, we study the shear response of the premelting film
upon sliding the wall with a constant lateral velocity 
$U=5$~m$/$s at T=262~K and $p=1$~atm for a period of 10~ns (Fig.~\ref{Fig2}-(a,b,c,d)). 
Similar calculations are performed for  $p=600$~atm (SI Appendix Fig. S6) and $p=1$~atm in the temperature range 230 to 266~K (SI Appendix Figs. S3 and S7).

Averaging the velocity components in the $x$ direction parallel to the
slider, we obtain the hydrodynamic flow profile $u(z)$ as a function of the
perpendicular distance to the wall.
For the hydrophobic wall (Fig.~\ref{Fig2}-(a)), $u(z)$ is
hardly distinguishable from thermal motion \cite{kannam13}. 
It appears as a noisy curve with
very small positive velocity that hardly attains 10\% of the total wall
velocity.
However, as the wall
hydrophilicity begins to increase (Fig.~\ref{Fig2}-(b,c,d)), a roughly linear flow profile 
appears that strongly resembles expectations from a model of simple Couette flow
with partial slip \cite{thompson90,barrat99,robbins00,kannam13,herrero19}. By visual inspection we can define $\sigma_1$ as the position close to the
wall adsorption peak where the approximately linear flow profile attains its
maximal velocity, $u_s=u(z=\sigma_1)$. Similarly, we define $\sigma_2$, close to the ice
adsorption peak, where the extrapolated flow profile vanishes, $u(z=\sigma_2)=0$ (See SI Appendix Fig.
S8, and Tables S1 and S2). 
Interestingly, we find $\sigma_2$ is at about one
molecular diameter away from the first ice bilayer, in agreement with reports of a small 
negative slip length for water flow past bulk ice \cite{louden17}.

If the hydrodynamics of the premelting film follows a model of Couette flow,
we expect the shear stress, $\tau$, should obey
$\tau=\eta\frac{u_s}{d_C}$,\cite{robbins00,persson00,barrat99}
where $u_s$ is the slip velocity, $u_s=u(z=\sigma_1)$,  $d_C$ is
the thickness of the region where the actual Couette flow takes place and $\eta$
is the bulk viscosity. To check this, we calculate the effective
hydrodynamic viscosity as $\eta_H=\tau d_C/u_s$, with
$\tau$ measured from the force exerted by the wall on the premelting film, 
$d_C=\sigma_2-\sigma_1$ and $u_s$ estimated by visual inspection of
the flow profile.  The results for $\eta_H$  
are shown in Fig.~\ref{Fig2}-(f), and appear
barely  10\% above the  viscosity of bulk water.
Alternatively, we can assume the premelting film behaves as bulk water
and obtain a hydrodynamic film thickness as $d_H=\eta u_s/\tau$, with
$\eta$ the bulk viscosity \cite{herrero19}. 
We checked that the 
value thus obtained agrees within $\pm 0.3$~nm 
with the  estimated Couette thickness, $d_C$. 
Now, using $d_H$, and $u_s$ 
we can obtain a synthetic Couette flow profile under the
assumption that $u(z)=u_s$ at $z=\sigma_1$ and vanishes at $z=\sigma_1+d_H$. The resulting model is
displayed in Fig.~\ref{Fig2}-(a,b,c,d) together with the actual velocity profiles
measured in the simulations.
A qualitative agreement is obvious for substrates
$\theta=107^\circ$ and $\theta=91^\circ$, and is almost as good as a linear
regression for the hydrophilic substrate with $\theta=50^\circ$
(similar good agreement is found also for $p=600$~atm,
c.f. SI Appendix Fig.S6 and $T=266$~K, SI Appendix Fig.S3 and S7). As a rule of thumb, we see that the
Couette flow is established between the wall and ice adsorption layers of
liquid--like water, so that the location of the hydrodynamic boundary conditions may be inferred with little cost from equilibrium simulations.

Put together, our results strongly support that the 
shear force of the slider on thin premelting films hardly one nanometer thick
may be described approximately by a very simple model of Couette flow with 
slip:
\begin{equation}
   \label{eq:model}
   \tau = \displaystyle \frac{\eta\, U}{h_H+b}
\end{equation} 
where $h_H=\sigma_1+d_H$ is the 
hydrodynamic film thickness, $b$ is a slip length and $\eta$ is an
effective viscosity similar to the viscosity of undercooled bulk water
(SI Appendix Fig. S8). 
We test the consistency of the model by performing additional simulations
at a smaller sliding velocity $U=0.5$~m$/$s and find that the calculated shear force is about 10 times smaller than
that measured at $U=5$~m/s, consistent with Eq.~(\ref{eq:model}) (See
SI Appendix Table S3).  

\begin{figure}[t]
   \centering
   \includegraphics[width=0.45\textwidth,trim=0cm 2cm 0cm 10cm,clip]{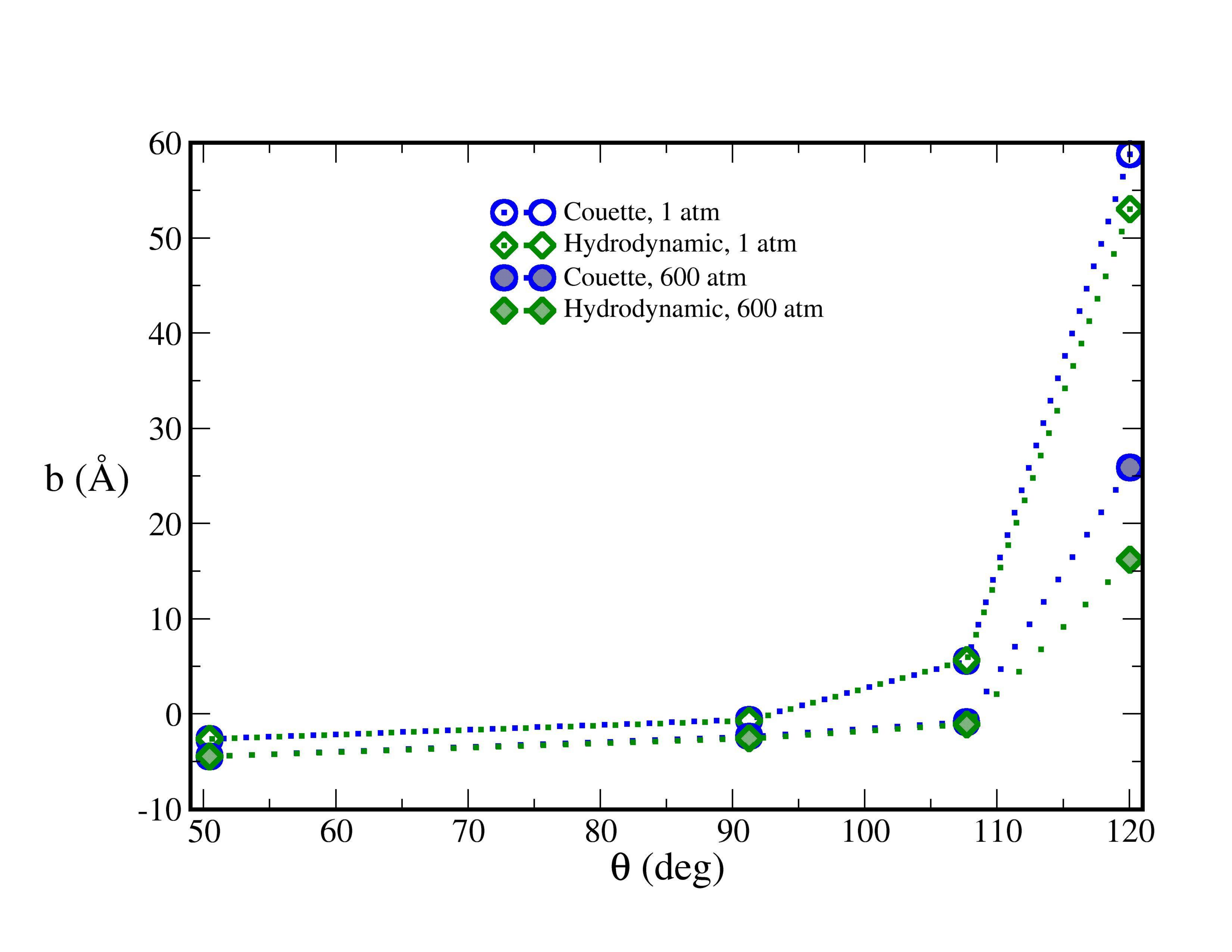}
   \caption{
	Slip and stick boundary conditions of premelted films. As wall hydrophilicity increases,
	the premelting film  dynamics evolves from large to negative slip.
	Blue circles are Couette slip lengths, $b_C$ 
	estimated from the average flow profile. Green diamonds are 
	hydrodynamic slip lengths, $b_H$ calculated from the shear force.
    } 
\label{Fslip}
\end{figure}

We can estimate a Couette slip length $b_C$ for use in Eq.~(\ref{eq:model})
directly from visual inspection of the
velocity profile of Fig.~\ref{Fig2}-(a,b,c,d), noticing that 
$U/(d_C+\sigma_1+b_C)=u_s/d_C$. 
A figure of the slip length as
a function of wall strength illustrates the large difference of frictional
behavior. For $\theta=120^\circ$, 
$b$ is about 5 times larger than the actual film thickness, consistent
with observations
of giant slip lengths in confined undercooled water and water at
hydrophobic substrates \cite{falk10,herrero20}.
As the wall strength increases, however,  the slip length
decreases fast and becomes negative for hydrophilic walls, illustrating
a large impact of wall interactions on the early stages of
ice friction (Fig.~\ref{Fslip}).

To further check the consistency of the model,
we can invoke the Navier slip boundary condition, which assumes
a shear stress proportional to the velocity drop at the hydrodynamic
boundary, i.e.
$\tau=\lambda (U - u_s)$, where $\lambda=\eta/b$ is the
interfacial friction coefficient \cite{barrat99,hansen11,kannam13,herrero19}. Combining these equations, we estimate a 
hydrodynamic slip length $b_H=(U-u_s)\eta/\tau$ by using the shear stress
obtained in the simulations, $u_s$ from the velocity profile and $\eta$ the 
viscosity of bulk water. The results in Fig.~\ref{Fslip} show remarkable good agreement
with the Couette slip length measured previously and attests to the accuracy 
of Eq.~(\ref{eq:model}) as a valid model for the shear stress of an atomically smooth
sliders on ice.

Since increasing the temperature or pressure increases the equilibrium film thickness, Eq.~(\ref{eq:model}) is expected to hold everywhere above T=262~K and $p=1$~atm. At lower temperature, however, Figs. S3 and S4 show that the premelting films  consist of at most two adsorption layers and no quasi-bulk region at all. This drives the system fully into the boundary friction regime. Surprisingly, the ice surface remains slippery both for hydrophobic and hydrophilic sliders. Indeed, using our results for the shear stress from the non-equilibrium simulations (SI Appendix Tables S3 and S4), and indentation hardness results for the applied load \cite{liefferink21}, we find order of magnitude agreement with experimental friction coefficients in all the temperature range studied (c.f. SI Appendix Text and Fig. S9). The origin of the small friction coefficients varies greatly with hydrophilicity, however. In the case of hydrophobic sliders, the diffusion coefficient of the adsorption layer remains always larger than the bulk diffusion coefficient (Fig.\ref{Fig2}-e), and allows for a  very large slip (Fig. S1 and S2), as suggested in Ref.\cite{weber18}. For hydrophilic sliders, on the contrary, shear is sufficiently large to melt one full bilayer in barely 0.5~ns due to frictional heating. As a result, an equilibrium premelting film consisting of one single adsorption layer attains nanometer thickness in less than 5~ns (c.f. Fig.S1 and Movie M3).Therefore, hydrophilic sliders develop a layer that is sufficiently thick to achieve lubrication, even at low temperature. This is visible in the close to linear velocity profiles observed in Fig.S7. However, for a nanometer thick film at the large sliding speeds $U=5$~m/s that we study, the shear rate $\dot{\gamma}=U/d$ attains the scale of $5\cdot 10^9$~s$^{-1}$. Above 262~K this is still small and the lubrication layer exhibits Newtonian flow. However, as the temperature decreases, this becomes well above the threshold of non-Newtonian flow \cite{ribeiro20}. Indeed, from the velocity profiles of Fig.S7 we find hydrodynamic viscosities that become up to two orders of magnitude smaller than the bulk viscosity at T=230~K. The shear rate dependent viscosities so obtained roughly follow the Eyring model of shear thinning (c.f. SI Appendix Text and Fig. S10).  Interestingly,  the hydrophilic sliders at low temperature display a small elastic deformation (c.f. Movie S3), but this accounts only for a few percent of the total shear stress (c.f. SI Appendix Text).

\section*{Discussion}

In practice, ice friction is a multiscale problem, and both the slider and ice are atomically rough \cite{vanossi13,muser17}. As a result, it is thought that most of the slider's load
is supported locally in high--pressure zones \cite{bowden53,colbeck88,lozowski13,weber18,liefferink21,lever21b}. 
In conventional applications, low viscosity liquids such as water behave as
very poor lubricants, because the large pressure between asperities  squeezes
out the lubrication
film, resulting in bare contact friction
\cite{colbeck88,robbins00,persson00,liefferink21,lever21b}. This argument was
used recently by Canale et al.\cite{canale19} and Bonn et
al.\cite{weber18,liefferink21} to put into question the role
of premelting mediated lubrication in ice friction. Our results show
ice does not behave as an ordinary inert substrate. Increasing the pressure
drives it closer to the melting line, and leads to an increase of the
equilibrium
premelting thickness (c.f. Fig.\ref{Fthickness}).
Of course, a high contact pressure will conspire to squeeze out the
lubrication film \cite{pittenger01}.  By Le Chatelier's principle, however, ice will melt
in order to restore the equilibrium thickness \cite{dash99,wettlaufer99b,sibley21}. Due to this self-healing
property of the premelting film, we
expect  lubrication will be enhanced at high pressure.

Based on these considerations, and the model of Eq.~(\ref{eq:model}), we find that the 
friction coefficient of an
atomically smooth region on the ice/slider system should obey 
$\mu_f = \frac{\eta(\dot{\gamma}) U}{(h_H+b)p}$, where $p$ is the pressure of the
high pressure zone \cite{colbeck88,lozowski13}.

  At  $T=262$~K, moderate sliding speeds of
the order of mm/s, assumed film thicknesses of ca.~1~nm and zero slip, this
gives already a small friction coefficients at a pressure of one atm, ca. $\mu_f=0.1$.  Considering instead that the sliding contact is exercised at the indentation hardness limit of ca. $p=100$-$1000$~atm \cite{tusima11,lozowski13,weber18,liefferink21}, 
we obtain friction coefficients two to three orders of magnitude smaller.
Increasing the sliding velocity to the m/s range yields higher estimates of
order $\mu_f\approx 1$, well away from experimental measurements. However, at
such ranges, friction inputs heat at a large rate of $\eta U^2/(h_H+b)$ ca.
35-200 MJ/m$^2$s, which is sufficient to melt roughly one full bilayer in the scale  of nanoseconds.
Unless heat is dissipated within the slider at a very fast rate,
this will result in a large increase of the film thickness 
within a few hundred nanoseconds, as assumed tacitly in current theories of
frictional heating \cite{colbeck88,lozowski13}.
Indeed, we can see in  our 10~ns sliding simulations clear evidence of bilayer melting at temperatures as low as 230~K, despite the use of a thermostat  (c.f. Movie S3).

Our model is also consistent with the temperature dependence observed for the
friction coefficient \cite{bowden53,budnevich94,tusima11,weber18,liefferink21}.
For hydrophobic walls at low temperature, the slip length becomes very
large \cite{herrero20}, and  the lubrication model with slip yields $\mu_f =
\lambda U/p$, where $\lambda$ is the interfacial friction \cite{barrat99}.  For
undercooled water,  $\lambda(T)$ follows an Anti-Arrhenius
behavior\cite{herrero20} that is consistent with a large increase of $\mu_f$ with
decreasing temperature \cite{budnevich94,tusima11,weber18,liefferink21}. The
result $\mu_f = \lambda U/p$ also explains the velocity strengthening observed 
for the
friction coefficient at slow velocities \cite{tusima11,schulson12,liefferink21}. At larger sliding
velocities, shear thining becomes significant. This leads to a friction coefficient with a weaker, logarithmic dependence on the
	sliding speed, $\mu_f \propto \ln(U)/p$ (c.f. SI Appendix Text). When the pressure is equated to
	the indentation hardness, which increases faster than $\ln(U)$ \cite{liefferink21}, this  results in the velocity
	weakening of the friction coefficient found in experiments  \cite{bowden53,budnevich94,tusima11,schulson12,liefferink21}.

Overall, we find our results lend strong support to the hypothesis of
lubricated ice friction that has
been put into question in recent experiments based on milimeter scale probes
\cite{weber18,canale19,liefferink21}.
Contrary to findings by Canale et al.\cite{canale19} for corrugated probes our
results suggest that at atomically smooth contacts, the effective viscosity
 $\eta$ is a meaningfull and
well defined
parameter, on the order of the bulk viscosity, which exhibits shear thinning at low temperature. 
Unlike suggestions by Weber et al. \cite{weber18}, the dependence of 
ice friction on substrate's slip length $b$ shows that the sliding dynamics
cannot be generally correlated with the properties of premelting films at the
ice/vapor interface, except for highly hydrophobic sliders.  

We emphasize, however that
our results describe the frictional behavior at
atomically smooth contacts. At a larger scale, both the slider and ice exhibit
microscale roughness, and the total load of the slider is supported by a small
amount of asperities.  Accordingly, the friction coefficient is
not only given by the shear at the smooth contacts, but also,
by the indentation hardness of ice, which sets the fractional area 
supporting the slider's load \cite{tusima11,lozowski13,weber18,liefferink21}. 
At a 
larger scale, the formation of a slurry of
water and ice could 
result in a very complex visco-elastic
response \cite{canale19,lever21b}.

In summary, we have shown that a very small extent of interfacial premelting in ice provides a lubricating quasi-liquid layer that can be described close
to quantitatively by a model of bulk Couette flow with slip. The premelted layer
can further grow by compression and frictional heating. Our results reconcile
the long--standing controversy on the origin of ice slipperiness and show that  equilibrium premelting, pressure melting and frictional
heating operate simultaneously.







\section*{Methods}

\subsection*{Computer simulations}
Molecular Dynamics simulations on the Np$_z$AT ensemble
were performed using LAMMPS \cite{lammps22}. Trajectories were evolved with the velocity-Verlet algorithm, 
with a time step of $2$ fs. 
Bonds and bond angles were constrained by the use of the SHAKE algorithm.
The temperature was set using the velocity rescale algorithm with damping factor
$\tau=0.2$ ps \cite{bussi07}. The pressure was set by applying a constant normal
force $F_z=\pm p_z A/N_w$ directed in the direction of bulk ice on each of the
$N_w$
wall atoms \cite{herrero19,dilecce20,herrero20}. 
This avoids perturbation of the dynamics that is usual in conventional 
barostats. To prevent strain effects in the parallel direction, the lattice
parameters are set to the equilibrium value at pressure $p_z$.
Shear was imposed on ice by moving the walls
tangentially in the $x$ direction at constant sliding speed (with opposite
direction on each wall) \cite{barrat99,robbins00,herrero19,ribeiro21}.
All Lennard-Jones interactions were truncated beyond 1~nm. Electrostatic
interactions were evaluated with a particle-particle particle-mesh solver. The
charge structure factors were evaluated with a grid spacing of $1$ \AA~and a
fourth order interpolation scheme. It resulted in $36 \times 32 \times 120$ ($72
\times 64 \times 120$ during Couette flow simulations) vectors in the $x, y, z$
reciprocal directions, respectively. In the Non-Equilibrium simulations,
equilibrated bulk systems were replicated by a factor of two on each of the
parallel directions to gather sufficient statistics for the flow profile, and
the thermostat was set only in the directions perpendicular to the flow, using a
damping factor $\tau=1$~ps. Shear was imposed on ice by moving the walls
tangentially in the $x$ direction at constant sliding speed for 10~ns (with
opposite direction on each wall). The averages were collected over 10
independent runs. Further details on the simulation setup and analysis may be found in the SI Appendix Methods and Figs. S11-S15. 

\subsection*{Data Availability}

All the study data are included in the article and the supporting information.

\acknowledgments{
We would like to thank 
Eva G. Noya for providing us with a code for the CHILL+ order parameter
and G. de Vilhena for careful reading of the manuscript. We would also like to thank 
Steve Plimpton for help with LAMMPS.   We acknowledge funding from the Spanish
Agencia Estatal de Investigaci\'on under research grant PIP2020-115722GB-C21. 
PL also thanks Ministerio de Ciencia e Innovacion for financial support under 
a Juan de la Cierva fellowship FJC2019-041329-I.

}



\onecolumngrid

\setcounter{page}{1}
\setcounter{table}{0}
\setcounter{figure}{0}
\pagenumbering{arabic}

{\centering
   {\large Supporting Information for}

   {\Large Ice friction at the nanoscale
	   \\ by \\}
	   {\large {\L}ukasz Baran$^a$, Pablo Llombart$^b$, Wojciech
	   R{\.z}ysko$^a$ and  Luis G. MacDowell}
}

\section*{Choice of wall model}

	Notice we have chosen a generic model substrate with an FCC crystalline
	order. This choice is a matter of convenience, since the FCC lattice in the (111) 
	direction can be made to match perfectly the hexagonal ice lattice in the basal 
	direction (c.f. SI Appendix Methods). However, many metals such as Pt, Pd, and Ru  exhibit close 
	match with ice, c.f. Ref.(41), 
	so the lattice geometry employed here is not unrealistic. 
	
	In practice, we set the value of the Lennard-Jones range parameter,
	$\sigma_{wo}=3.1668$~{\AA},  equal to that of TIP4P/Ice water for convenience.  This is somewhat larger than the usual values for metals (c.f. 2.55~{\AA} for Cu and Fe),  and somewhat smaller  than those of CH2 and CH3 groups in united models of alkanes 	(ca. 3.9~{\AA} and 3.7~{\AA} for the OPLS or TraPPE models, SI Appendix Ref.\cite{jorgensen84,martin98}). 
	Therefore, our substrate is
	reasonable but does not particularly match any specific choice of material.
	However, by tuning the LJ energy parameter, $\epsilon_{wo}$, we tune the contact angles from 
	50 to 120 degrees (see the last section of this Appendix).
	For inorganic materials this includes  contact angles ranging from 
	low-energy metals to monolayer graphene. For plastic materials, this ranges from
	hydrophilic plastics such as Nylon-6 to highly hydrophobic ones such as PTFE.
	This is a large range of all possible contact angles relevant to a flat substrate, so we believe that, despite the need	to make some specific choice of the model, we are exploring a large set of	conceivable outcomes.
	
	Indeed, the main role of the wall-water interactions is to tune the relevant hydrodynamic boundary conditions via the slip length. According to a large body of theoretical and computer simulation results, the slip length of crystalline surfaces
is a universal function  of the variable $\xi=S(q^*) <F_x^2>$, where $S(q^*)$ is the
in-plane structure factor of the first adsorption layer at wall's smallest wave-vector; while  $<F_x^2>$  is the average lateral
wall-fluid squared force, c.f. Ref.(2,30,47,48). 
 Therefore, we believe that changing one single wall parameter in a way that spans $\xi$  over its full range is covering most of the relevant physics.
	
Another concern with the choice of the wall model is the commensurability with the ice lattice. In principle,	this could have two undesirable effects:
\protect\paragraph{It greatly stabilizes the wall/ice interface, so that it could  inhibit premelting.}
		In practice, our simulations show that compressing the ice/vapor interface results
		in a stable premelted layer even with a  perfectly commensurate wall.
		Using an incommensurate wall will enhance this effect. Whence,
		the stabilization of the premelting layer that we report remains robust whether
		the substrate is commensurate or not.  
\protect\paragraph{It removes strain on the ice lattice.} If ice is in direct contact with the wall, the perfect match will remove spuriously the strain that would result otherwise. However, we have seen above that approaching the solid substrate to the ice/vapor interface results in the stabilization of the premelting layer, and more likely so if the wall were incommensurate with the ice lattice. Therefore, the possible strain will be relaxed within the liquid layer, and strain effects are therefore not a concern.

	\section*{Comparison with experimental friction coefficients}
	
	In principle, our results for the shear stress displayed in Tables
	\ref{stress}--\ref{stress_temps} can be used to estimate friction coefficients. However, a
	quantitative comparison of such results with experimental friction coefficients
	is not possible.  The reason is that friction is a multiscale 
	problem and the actual coefficients that are measured depend not only
	on the substrate's properties, but also on the ice and slider surface
	preparation and roughness, the slider's length and geometry, as well as
	the length of the track, the time of sliding and past history, c.f. Ref.(22-24) and
	SI Appendix Ref.\cite{muser03}.
	 Particularly, it is believed
	that most of the slider's load is supported in small asperities, so that
	the real area of contact is actually unknown, and the pressure on the
	asperities cannot be measured directly.  Although experiments can not measure directly the
	pressure on asperities, it is usually assumed  that its value is given
	by the indentation hardness, which depends both on temperature and penetration
	speed, c.f. Ref.(13).
	At a sliding speed of 5~m/s that we study, Liefferink et al.
	estimate a penetration speed of about 0.05~m/s,  which corresponds to a linear
	estimated hardness of $p_H = 440 - 2.6\times (T-272)$~MPa, in order of magnitude
	agreement with computer simulations of indentation hardness for the TIP4P/Ice
	model at 1~m/s  (c.f. SI Appendix Ref.\cite{santosflorez20}). At the range of
	temperatures studied, this corresponds to pressures above the melting point,
	whence, an alternative estimate that is plausible if the melting rate is
	faster than the penetration speed is to assume the asperities withstand
	a pressure similar to the melting pressure, i.e. $p_m = -13.5\times(T-272)$~MPa. 
	
	Calculations of the shear stress in our simulation have been performed for
	significantly smaller applied pressure. However, our model of shear stress, Eq.(1) 
	of the main text, exhibits only a weak dependence on
	pressure, via the shear viscosity, so we can obtain an order of magnitude comparison of 
	friction coefficients by dividing the simulated shear stress by the estimated 
	pressure on the asperities. 
	
	Figure \ref{mu_v_exp} displays the friction
	coefficients estimated from  $\mu=\tau/p_i$, with $p_i$ either equal to
	the indentation hardness, $p_H$ or the melting pressure, $p_m$, for hydrophobic and hydrophilic
	sliders. The results displayed, bracket experimental measurements of friction
	coefficients at similar sliding speeds, and show either increasing or
	decreasing friction coefficients with T, which are the two possible outcomes
	encountered  in experiments depending on the material, as shown in the figure.

	\section*{Shear Thinning}

Using the model of Eq.(1) in the main text, which assumes bulk Newtonian viscosity,
the prediction of the hydrodynamic film thickness that is obtained, $d_H=\eta u_s/\tau$, is orders of magnitude too large at 230 and 240~K, and about two times larger at 250~K. This results in predicted flow profiles that have an extremely low decay, as shown in Figure \ref{FShear_Density_f40}. Clearly, the correct shear stress can only be predicted from Eq.(1) of the main text if we assume a much smaller effective viscosity. 

As noted in the main text, shear rate dependent viscosities can occur  at low temperature for shear rates larger than a temperature dependent threshold value, a phenomenon known as shear thinnig, c.f. Ref.(2) 
and SI Appendix Ref.\cite{muser03,spikes14}. Recently, de Almeida Ribeiro and de Koning studied the rheology of supercooled water with the TIP4P/Ice model and found  significant shear thinning below 250~K for shear rates above $10^9$s$^{-1}$, which is roughly the shear rate attained in our simulations, c.f.  Ref.(50). 
Therefore, we expect the small effective viscosities required to describe the actual shear stress in the system to result from shear thinning.

According to the Eyring theory of shear thinning (SI Appendix Ref.\cite{spikes14}), the shear viscosity of a fluid obeys the following equation:
\begin{equation}
 \frac{\eta(\dot{\gamma})}{\eta(0)} =  \frac{\tau_0}{\dot{\gamma}\eta(0)}\sinh^{-1}\left(\frac{\dot{\gamma}\eta(0)}{\tau_0}\right)
\end{equation}
where $\eta(0)$ stands here for  the bulk Newtonian viscosity, and $\tau_0$ is a threshold shear stress above which shear thinning becomes significant. The shear threshold depends on the temperature as 
$\tau_0 = \frac{k_BT}{v_a}$, where $k_B$ is Boltzmann's constant and $v_a$ is an activation volume in the order of the molecular volume.

As shown in Fig.~\ref{shear_thinning}, this single parameter model describes in reasonable agreement the shear rate dependent hydrodynamic viscosities obtained from our simulations as $\eta(\dot{\gamma})=\tau d_C/u_s$. Moreover, the fit provides $v_a=1.5\cdot 10^{-28}$~m$^3$, which corresponds to an effective radius $\sim 3\cdot 10^{-10}$~m of the same order of magnitude as the molecular radius, in agreement with expectation from the Eyring model. 

Since $\sinh^{-1}(x)\approx \ln(x)$ for large $x$, the Eyring model
predicts a shear stress $\tau=\tau_0\ln(\dot{\gamma}\eta(0)/\tau_0)$ which, in view of
$\dot{\gamma}\approx U/d_C$, yields a logarithmic dependence of shear stress with sliding velocity, as found in Ref.(13), but see also SI Appendix Ref.\cite{muser03}. 
It follows that the
friction coefficient, $\mu_f = \tau / p$, becomes of the order $\ln(U) / p$, as explained in the main text.

	\section*{Contribution of Elasticity}

	For the hydrophobic slider at temperatures below 250~K, some amount of
	strain is observable in the simulation snapshots (c.f. Movie S3).
	Inspection of the molecular configurations shows a lateral displacement of
	about one lattice position in the direction of the slider. This
	corresponds to ca. $\frac{1}{4}b$, where $b$ is the crystal unit length in
	the sliding direction. This deformation is propagated over roughly 24 unit
	cells in the z-direction, of size $24 c$ (with $b\approx 7.9$~\AA$\,$ and
	$c\approx 7.4$~\AA). This yields a small shear strain of
	$\gamma = \frac{1}{96} \frac{b}{c} \approx 0.01$. 
	The shear modulus of ice in the temperature range studied is about $G=3$~MPa, according to  Ref.(11).
	 Therefore, the shear stress expected from the elastic
	deformation is $\tau = G\gamma=3\cdot10^{-2}$~MPa, which is just a small
	fraction of the full shear stress measured in the simulations (c.f. Table
	   \ref{stress_temps}). Interestingly, a small elastic response has also been
	   measured in recent experiments by Canale et al.(14). 

\section*{Methods}

\subsection*{Model and setup}

Water was modeled with the TIP4P/Ice force field, Ref. (33). 
Wall-water
interactions were implemented with a force-shifted Lennard-Jones potential between
wall atoms and water oxygens. For the wall-oxygen interactions, we chose
$\sigma_{wo}$ equal to $\sigma_{oo}$ in the TIP4P/Ice model, and tuned the
wall-oxygen $\epsilon_{wo}$ to $f\epsilon_{oo}$, with $f=\{1,2,3,4\}$. All
dispersion interactions were truncated at 1~nm. An initial configuration of ice
Ih oriented along the basal direction was prepared by replicating a
pseudo-orthorombic unit cells of size $(2\times a)\times b\times c$, with 16 water molecules each.
The bulk slab consisted of an arrangement of $4\times 4\times 15$ such cells, prepared so as to
leave a half terminated bilayer exposed on the surface. {Notice that
in view of the limited amount of premelting observed, the simulation cells are unnecessarily large in the perpendicular direction, since the stress tensor components decay exponentially fast}. A random hydrogen bond network with total zero dipole moment was created following Ref.(46). 
The wall consisted of a stack of three close--packed planes in a
face-centered cubic (FCC) arrangement oriented along the (111) direction. A perfect
match of the wall with the ice basal face is achieved by
choosing a wall unit cell of size $a$, $b=\sqrt{3}a$ and
$c=\sqrt{6}a$. For each temperature and pressure, we slightly rescaled the wall and ice
unit cells to the corresponding equilibrated ice lattice.
The ice slab was then sandwiched between the walls, and
the whole system was placed in a simulation box under periodic boundary
conditions, leaving a gap between the walls across the boundary
conditions. 

In order to save computational time, we do not put the walls in contact with a fully equilibrated ice/vapor interface at the relevant temperature. Instead, we have found that arranging the bulk ice lattice such that the external layer of the slab is a half terminated bilayer, instead of a full bilayer, does well the job. We illustrate this in Fig.\ref{comparison_total_densityp}, which shows that the final equilibrated density profile is exactly the same, whether the simulation starts from the half terminated bilayer of a pre--equilibrated ice/vapor interface. Care must be taken when one puts directly the fully terminated bilayer in contact with the wall. In this case, the perfect wall match can stabilize an ice slab with no premelting layer whatsoever for a long time.

	\subsection*{Analysis}
	
	The CHILL+ order parameter was employed to label water molecules in 
	liquid-like and solid-like categories. Molecules labeled as bulk crystalline or
	interfacial crystalline (mainly Ih, Ic and interfacial Ih ice) were assigned as solid-like, 
	and the remaining molecules were assigned as liquid-like. Parallel diffusion
	coefficients of the interfacial layer were calculated by monitoring the mean
	squared displacement of liquid--like molecules within the assigned interfacial
	region. Bulk transport coefficients were calculated using equilibrium Molecular Dynamics
	in the NVT ensemble. Diffusion coefficients were evaluated from mean square
	displacements and shear viscosity from Green-Kubo relations. Velocity profiles were smoothed using an unweighted moving average with spatial extent of 4~\AA.

\subsection*{Implementation of the CHILL+ order parameter}

In order to characterize the premelting film, each water molecule in the system is labeled according to the CHILL+ algorithm, Ref.(37). 
This algorithm allows one to identify
ice allotropes as well as clathrates and interfacial ice Ih.
To determine the amount of liquid-like and solid-like molecules we proceed as follows. First, we have identified the largest ice cluster
and labeled all the molecules as solid-like molecules. In our case, this lumped regular ice Ih, interfacial ice Ih, and ice Ic into the solid-like category.
We have found that  CHILL+ treats some of the water molecules
as mislabeled due to not fulfilling any of the criteria specified in Ref.(37).
To assign them in either the solid or liquid group, we visualized the system and found that they always appeared within the liquid layer. Therefore, they are also included in the group of liquid-like molecules. Figure~\ref{Fdensity_il} displays results for the density of solid and liquid-like molecules. The plot shows that solid-like molecules penetrate slightly within the premelting layer, with some small oscillatory behavior found at high pressure. The presence of small ice patches could result in an additional viscoelastic response that is not taken into account in our model of Couette flow with slip. The consistency of the model indicates that viscoelastic contributions must be small in the regime studied in this work.

\subsection*{Measure of shear stress and test of barostat}

The ice sample was compressed by imposing a force $F_z=\pm P A/N_w$ perpendicular to the interface in the direction of bulk ice on each of the wall atoms, c.f. 
Ref.(29,30,38) and 
SI Appendix Ref.\cite{heyes11,marchio18,amabili19}.

The pressure exerted by the wall on the fluid must be balanced by the corresponding force exerted by the fluid on the wall. Whence,
it must follow (c.f. SI Appendix Ref.\cite{henderson92b}):
\begin{equation}
p_z=- \frac{1}{A}\int\frac{d\phi}{dz}\rho({\bf r}) d{\bf r}
\label{internal}
\end{equation}
\noindent where $\phi$ is the potential energy between wall atoms and water molecules and $\rho({\bf r})$ is the density profile. For an atomically resolved density field, this amounts to the calculation of: 
\begin{equation}
p_z=\frac{1}{A}\sum_i f_z({\bf r}_i)
\label{internal1}
\end{equation}
where $f_z$ is the $z$ component of the force exerted by the wall on a water
molecule at ${\bf r}_i$, and the sum runs over all water
molecules within the cutoff distance from the wall. We use this result as a test
of consistency for the barostat and find excellent performance as illustrated in Table ~\ref{stress} and Table ~\ref{stress_temps}.

Likewise, the shear stress $\tau$ is calculated from the total force exerted by the wall atoms on the water molecules in the direction of the slider. Whence, $\tau=p_x$, and
\begin{equation}
p_x = \frac{1}{A}\sum_i f_x({\bf r}_i)
\end{equation} 
with the sum convention as explained in the previous equation.

\subsection*{Estimation of contact angles}

The wettability of the substrates is characterized by the
contact angle of water sessile droplets at ambient temperature, T=298~K.
Unfortunately, measuring contact angles is not a trivial matter, because
they exhibit a very large system size dependence due to line tension effects
(SI Appendix Ref.\cite{macdowell02,vazquez09}).  In order to estimate the macroscopic  
contact angle, $\theta_{\infty}$, we calculated equilibrium contact angles for small droplets
with 2304 and 5120 water molecules, and extrapolated to the infinite droplet
size as (SI Appendix Ref.\cite{pethica77}):
\begin{equation}
\cos(\theta) = \cos(\theta_{\infty}) - \frac{\tau}{\gamma_{lv}} \times
\left ( \frac{1}{R \sin(\theta)} \right )
\end{equation} 
where $\tau$ is the line tension and $\gamma_{lv}$ is the liquid-vapor surface
tension. 

Based on the values of the contact angles extracted from these simulations (cf. Table~\ref{tab6})
we have extrapolated the curves shown in Figure~\ref{cost} to $1/r=0$ where $r=R\sin(\theta)$.
In this notation, the $R$ and $r$ refers to the radius of an auxiliary sphere and the radius of the base of the drop, respectively.
Estimated values of the contact angles in the infinite droplet size are:

\begin{description}
	\item{f=1} $\theta_{\infty}=120^\circ$
	\item{f=2} $\theta_{\infty}=107^\circ$
	\item{f=3} $\theta_{\infty}=91^\circ$
	\item{f=4} $\theta_{\infty}=50^\circ$
\end{description}


To measure the equilibrium contact angles of small spherical droplets we first
simulated samples of a bulk liquid water in the $NpT$ ensemble for 15 ns at $T=298$ K and $p=1$ atm. 
Then, the water parcel was placed on the solid surface and allowed to
equilibrate. The relaxation of the initial configuration is very slow.
Simulations were carried out over  60 to 100 ns at $T=298$ K in $NVT$ ensemble in
order to attain meaningful averages.
From the trajectories collected over the final 10 ns of the simulations average
density profiles $\rho(x,y)$ have been evaluated following the procedure
described elsewhere  (SI Appendix Ref.\cite{wloch16}).
Briefly, the droplet has been divided into cylindrical slabs of the same width
and within each, rectangular prisms have been further defined
in which the average density has been calculated.
The interfacial points of each slab have been located as
$\rho(x,y)=(\rho_{liquid}-\rho_{vapor})/2$.
The resulting density profile is fitted to a spherical cap of radius $R$,
centered at position $z_0$ (SI Appendix Ref.\cite{wloch16}).  In order to calculate the values of the contact angles, 
the  curves have been extrapolated to the bottom of the droplet.
Estimated error in the contact angle value is $\pm 2^\circ$ depending 
on the fitting, i.e. whether we fit entire drop or up to the $3/4$ of its height.

Similarly to all previous simulations, the charge structure
factors were evaluated with a grid spacing of $1$ \AA~and the
fourth order interpolation scheme. It resulted in the $54 \times
48 \times 30$  ($72 \times 64 \times 40$)
vectors in the $x, y, z$ reciprocal directions, respectively,
for 2304 (5120) water molecules.
In the case of the simulations for droplets sitting on a solid
surfaces, the number of vectors were equal to
$150 \times 144 \times 100$ ($200 \times 180 \times 120$) in the
$x, y, z$ reciprocal directions,
respectively, for 2304 (5120) water molecules.


\clearpage

\newpage

\begin{figure}[htb!]
	\centering
	\includegraphics[width=0.95\textwidth,trim=0 4cm 0 0, clip]{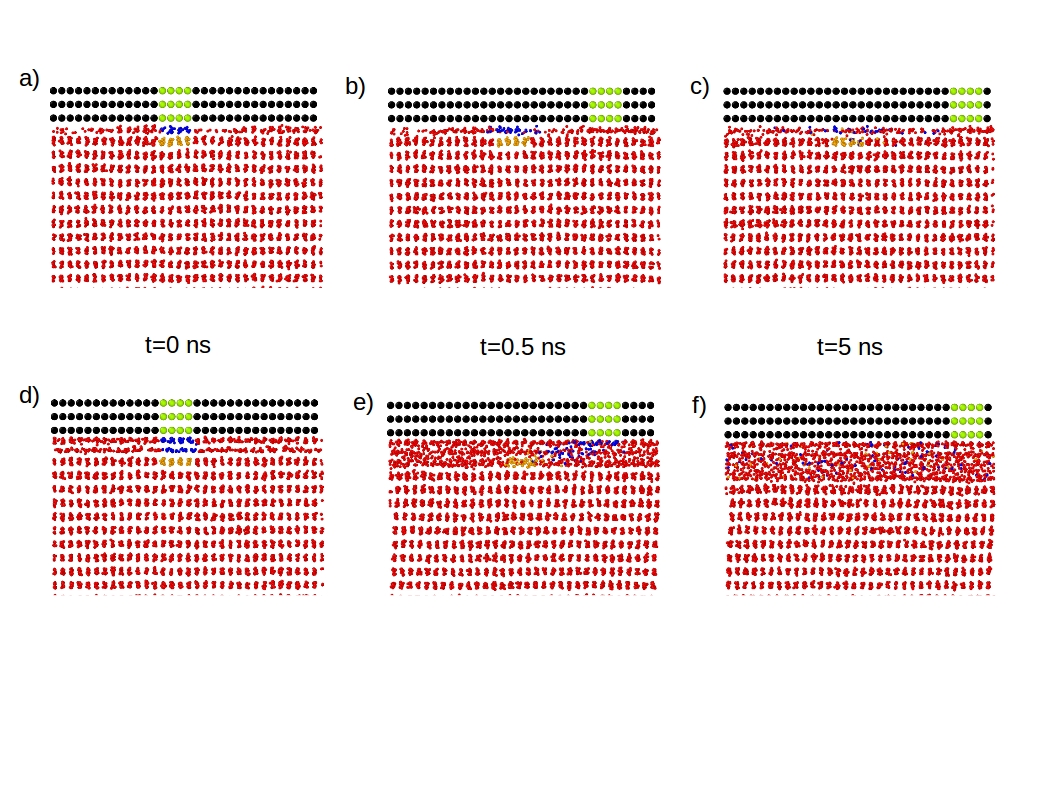}
	\caption{Time sequence of a shearing experiment at $T=230$~K. \textbf{(a-c)} A hydrophobic slider ($\theta=\fone^\circ$) barely one nanometer thick slips past the substrate. \text{(d-f)} A hydrophilic slider ($\theta=\ffour^\circ$) sticks to the substrate and exhibits significant frictional melting. Color codes as in Figure 1 of main text.
	} 
	\label{Fdisplay_T230}
\end{figure}

\newpage

\begin{figure}[htb!]
	\centering	
	\includegraphics[width=0.95\textwidth,trim=0 4cm 0 0, clip]{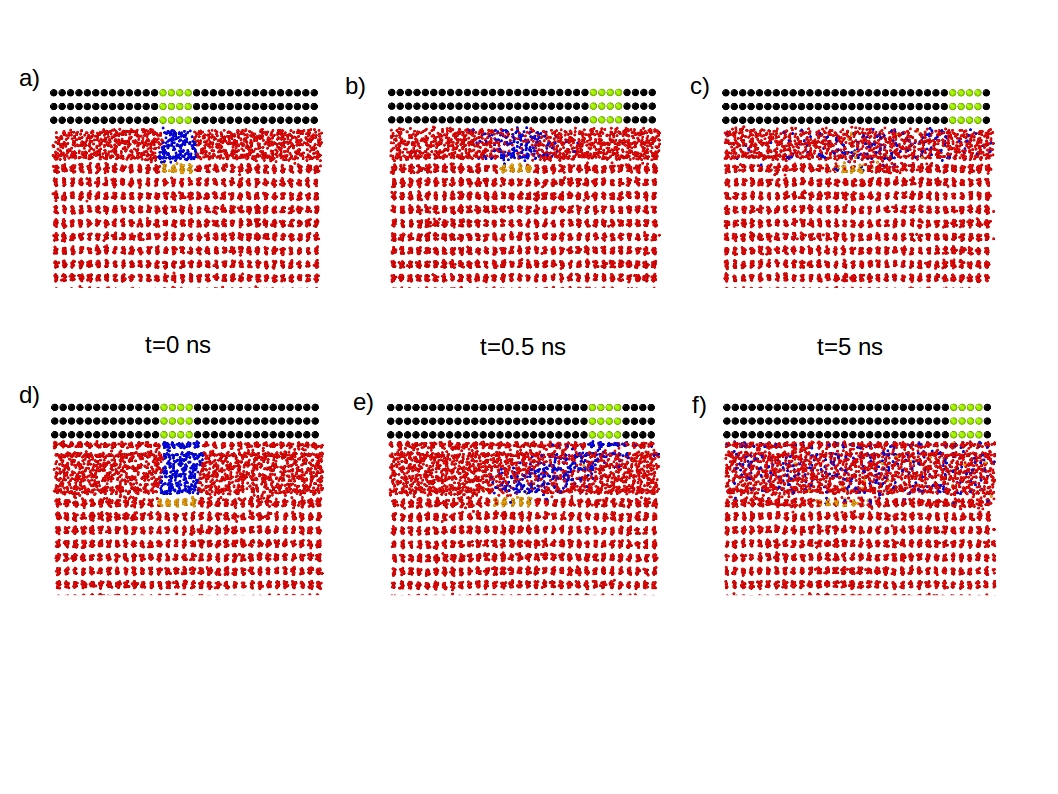}
	\caption{Time sequence of a shearing experiment at $T=266$~K. \textbf{(a-c)} The hydrophobic slider ($\theta=\fone^\circ$) slips. \text{(d-f)} The hydrophilic slider ($\theta=\ffour^\circ$) sticks. Color code as in Figure 1 of main text.
	} 
	\label{Fdisplay_T266}
\end{figure}

\begin{figure}[htb!]
	\centering
	\includegraphics[width=0.75\textwidth]{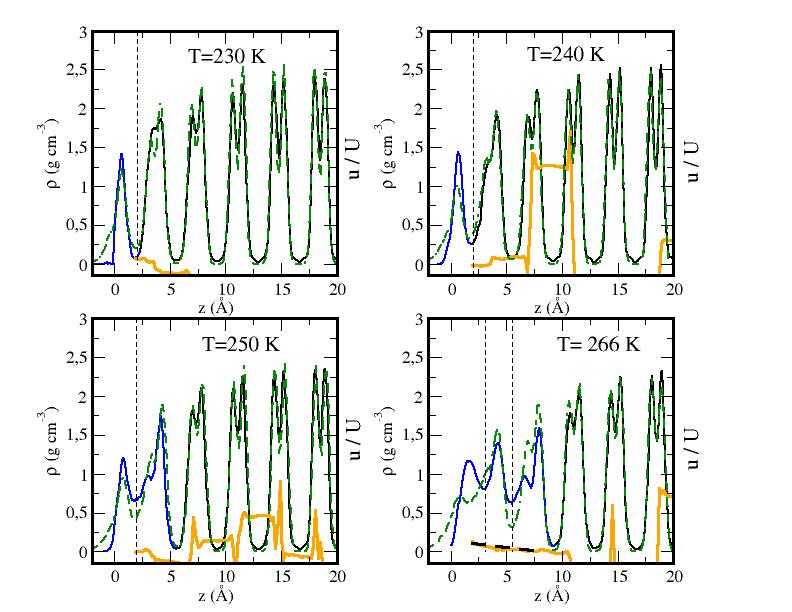}
	\caption{
		Structure and velocity flow of the premelted film.
		Results are shown for a hydrophobic slider ($\theta=120^\circ$) at $p=1$~atm in a range of temperatures.
		The equilibrium density profile in g~cm$^{-3}$ (left axis)
		is shown as a continuous line,
		with blue colour for the region where liquid-like water is the majority
		phase and black colour where ice is the majority phase. Green dashed lines
		describe the total density profile at the ice/air interface (results from SI Appendix Ref.\cite{llombart20}). Vertical black dashed lines serve to separate adsorption layers next to the wall and bulk ice. An additional line separates  a central quasi-bulk region where possible.
		The thick continuous lines display the velocity profile in units of
		the sliding velocity for a sliding experiment with $U=5$~m$/s$ during 10~ns (right axis). The dashed black line in the panel at T=266~K is the hydrodynamic flow profile predicted from the model of Eq. (1) in the main text. For lower temperatures the flow profile is too noisy to obtain any reliable information. 
	} 
	\label{FDensity_f10}
\end{figure}

\begin{figure}[htb!]
	\centering
	\includegraphics[width=0.75\textwidth]{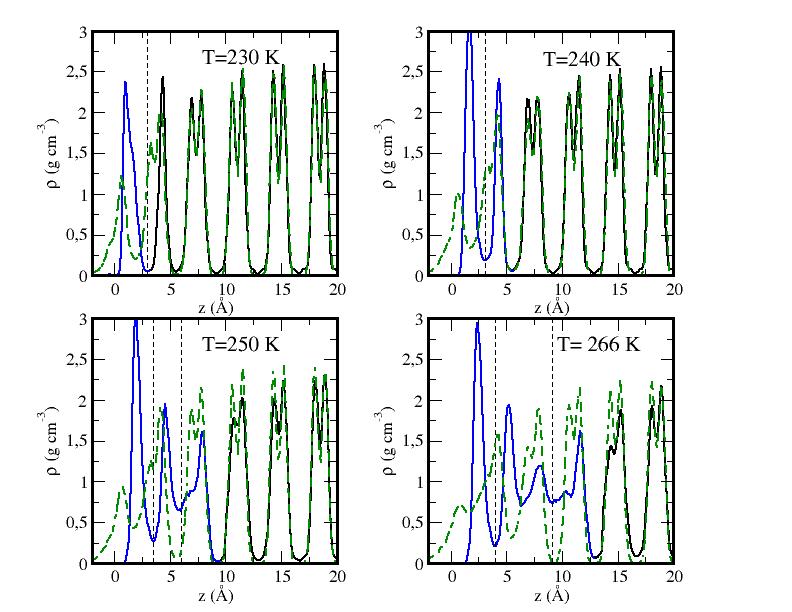}
	\caption{
		Structure and velocity flow of the premelted film.
		Results are shown for a hydrophilic slider ($\theta=50^\circ$) at $p=1$~atm in a range of temperatures.
		The equilibrium density profile in g~cm$^{-3}$ (left axis)
		is shown as a continuous line,
		with blue colour for the region where liquid-like water is the majority
		phase and black colour where ice is the majority phase. Green dashed lines
		describe the total density profile at the ice/air interface (results from SI Appendix Ref.\cite{llombart20}). Vertical black dashed lines serve to separate adsorption layers next to the wall and bulk ice. An additional line separates  a central quasi-bulk region where possible.
		Velocity profiles are not shown in this case because substantial frictional melting distorts the premelting film structure, but see Supplementary Figure \ref{FShear_Density_f40}.}
	\label{FDensity_f40}
\end{figure}

\begin{figure}[htb!]
   \centering
   \includegraphics[width=0.45\textwidth,trim=0 0 0 9cm,clip]{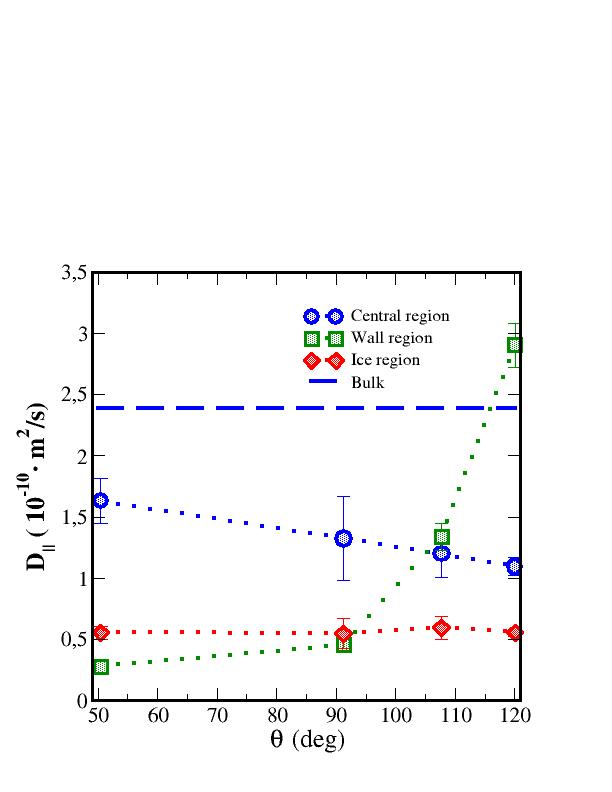}
   \includegraphics[width=0.45\textwidth,trim=0 0 0 9cm,clip]{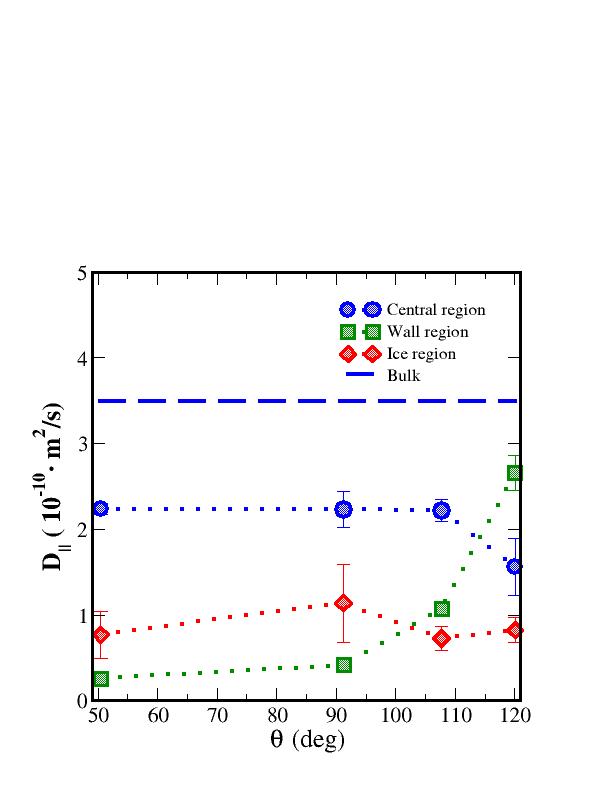}
   \caption{
	Quasi-Bulk like diffusivity of the premelting film.
	Results are shown for   $T=262$~K and $p=1$~atm (left) and $p=600$~atm (right).
	The thick dashed line
	stands for the bulk diffusion coefficient.
	The remaining lines correspond to parallel diffusion coefficients of
	liquid-like molecules in different regions of the premelting film: 
	Adsorption layer next to the
	wall (green squares); Adsorption layer next to ice (red diamonds);
	and quasi-bulk central region (blue circles). 
   } 
\label{Fdiffusion600}
\end{figure}

\begin{figure}[htb!]
	\centering
	\includegraphics[width=0.75\textwidth]{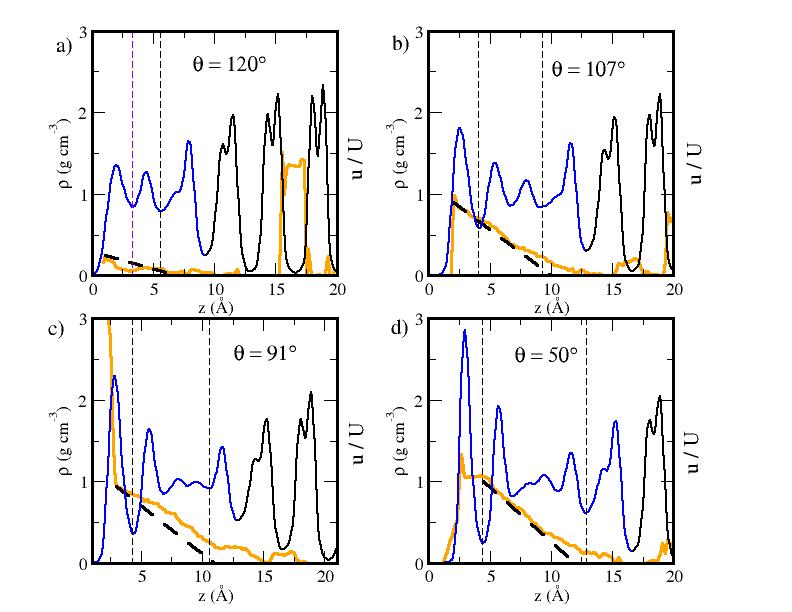}
	\caption{
		Structure and velocity flow of the premelted film
		during sliding.  Results are shown for a
		sliding experiment at $p=600$~atm, $T=262$~K and $U=5$~m$/$s during 10~ns.
		The equilibrium density profile in g~cm$^{-3}$ (left axis)
		is shown as a continuous line,
		with blue colour for the region where liquid-like water is the majority
		phase and black colour where ice is the majority phase.
		The premelting film is divided into two adsorption layers next to
		the wall and bulk ice and a central quasi-bulk region, as illustrated by
		vertical dashed lines.  The thick
		continuous lines display the velocity profile in units of
		the sliding velocity (right axis). The dashed black line is the
		hydrodynamic flow
		profile predicted from the model of Eq. (1) in the main text. Notice a smaller extent of
		slip compared to results at $p=1$~atm.  Panels correspond
		to different wall interactions  \textbf{(a)} Hydrophobic wall, with
		$\theta=120^\circ$ \textbf{(b)} $\theta=107^\circ$ \textbf{(c)} $\theta=91^\circ$ \textbf{(d)} Hydrophilic
		wall with $\theta=50^\circ$.
	} 
	\label{Fdensityp600}
\end{figure}

\begin{figure}[htb!]
	\centering
	\includegraphics[width=0.75\textwidth]{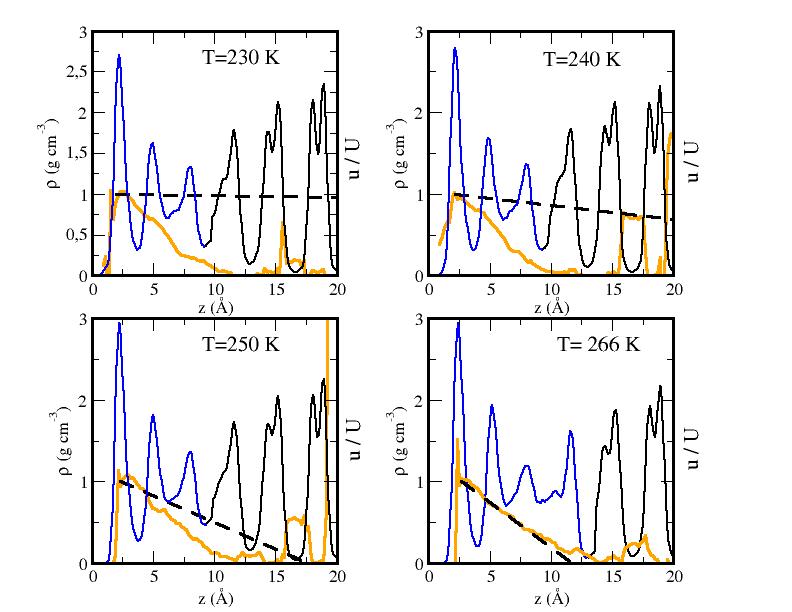}
	\caption{
		Structure and velocity flow of the premelted film
		during sliding.  Results are shown for a
		sliding experiment on the hydrophilic wall ($\theta=50^\circ$) at
		$p=1$~atm, and $U=5$~m$/$s during 10~ns. The density profile in g~cm$^{-3}$ (left axis)
		is shown as a continuous line,
		with blue colour for the region where liquid-like water is the majority
		phase and black colour where ice is the majority phase.
		Since these systems exhibit substantial
		frictional melting for temperatures below 262~K, the density profiles now correspond to the average obtained during the full 10~ns (c.f. Fig.\ref{FDensity_f40}) for the equilibrim density profiles of the same system).
		The thick orange
		lines display the velocity profile in units of
		the sliding velocity (right axis). The dashed black line is the
		hydrodynamic flow
		profile predicted from the model of Eq. (1) in the main text. Note complete failure of the model at the two lowest temperatures due to substantial shear thinning. Panels correspond
		to different temperatures as indicated.
	} 
	\label{FShear_Density_f40}
\end{figure}

\begin{figure}[htb!]
   \centering
   \includegraphics[width=0.85\textwidth]{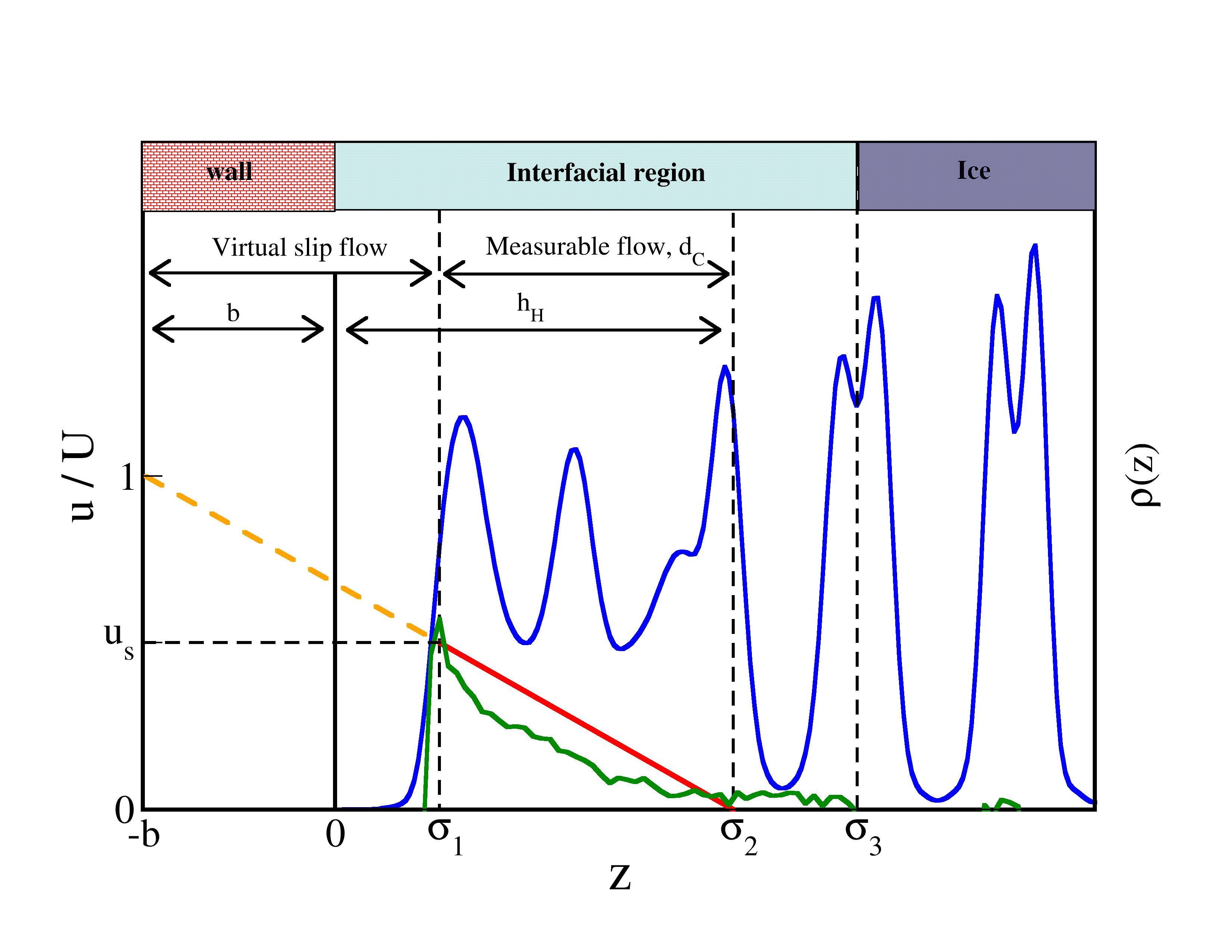}
   \caption{Sketch of hydrodynamic reference frame and boundary conditions. The
blue lines display a sample density profile. The green line is a sample
flow profile (average velocity in the direction of sliding as a function of
perpendicular distance to the wall).  The red line describes a hypothetical
Couette flow profile required to describe the actual shear force from
$\tau=\eta dU/dz$.
In order to describe the actual
flow in terms of a model of Couette flow with slip, $\tau=\eta U/(b_w+h_H)$, 
we place the origin of the 
hydrodynamic reference frame $z=0$ at the first layer of rigid wall atoms. 
The flow profile attains its maximum value, $u_s$, at position $\sigma_1$,
(close to the loci of the first adsorption peak of premelted water on the wall)
whereupon, it shows a sharp drop and then vanishes.
$\sigma_2$ is defined as a visual extrapolation of a straight line towards 
the bulk ice (close to the first adsorption peak of
premelted water on bulk ice). A small tail of the flow profile persists up to
the first ice bilayer, and vanishes at position $\sigma_3$. The difference
between $\sigma_2$ and $\sigma_3$ corresponds to a small negative slip for the
flow of water on ice.
The extrapolated flow profile within the bulk substrate attains the velocity of
the slider at a virtual position $z=-b$ which defines the wall slip length
in our hydrodynamic model.
    } 
\label{Fsketch}
\end{figure}
\clearpage

\newpage

	\begin{figure}[htb!]	
		\centering
		\includegraphics[width=0.95\textwidth]{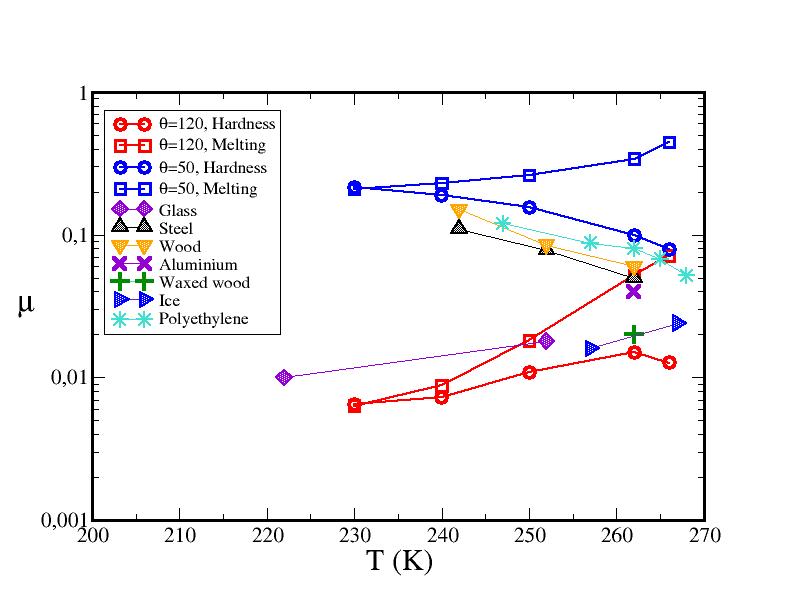}
		\caption{Consistency with experimental friction coefficients. Figure displays estimated friction
			coefficients at $U=5$~m/s from this work (empty symbols) with experimental data
			for different materials and sliding speed between 0.05 and 8~m/s (filled symbols).  Empty
			squares correspond to friction coefficients estimated from the indentation
			hardness as reported in Liefferink et al. (13). 
			Empty circles are
			obtained from the melting pressure. Experimental results are:
			Liefferink et al.(13), 
			glass sphere on ice, $U\approx 0.05$~m/s
			(violet diamonds);  Budnevich and Derjaguin (10),
			steel on ice,
			$U\approx 0.3$~m/s (grey triangles); Budnevich and Derjaguin (10), 
			wood on ice, $U\approx 0.3$~m/s (orange triangles); Bowden (9), 
			aluminium on ice, $U=5$~m/s
			(violet X); Bowden (9),
			waxed wood on ice, $U=5$~m/s (green cross).
			Oksanen and Keinonen (22),
			ice on ice (blue triangles), $U=3$~m/s;
			Stamboulides et al. (SI Appendix Ref.\cite{stamboulides12}) Ultra High Molecular Weight
			Polyethylene on ice  at $U=2$~m/s (turquoise starts).
		}
		\label{mu_v_exp}
	\end{figure}

\begin{figure}[htb!]
	\centering
	\includegraphics[width=0.95\textwidth]{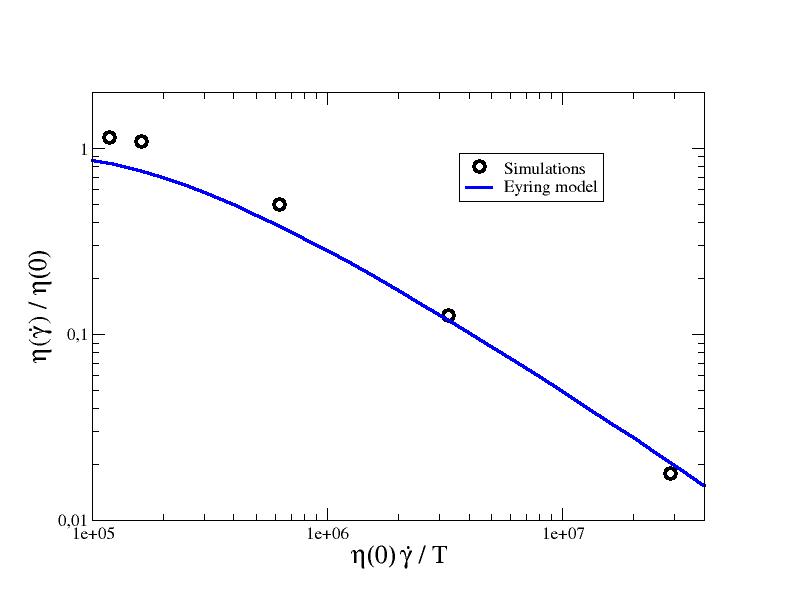}
	\caption{Shear thinning of premelting layers. The circles display the ratio of shear rate dependent viscosity to bulk Newtonian viscosity $\eta(\dot{\gamma})/\eta(0)$,  as a function of the single variable $\eta(0) \dot{\gamma} /T$. The full lines are a fit to the Eyring model of shear thinning.
}
\label{shear_thinning}
\end{figure}

\newpage

\begin{figure}[htb!]	
	\centering
	\includegraphics[width=0.95\textwidth]{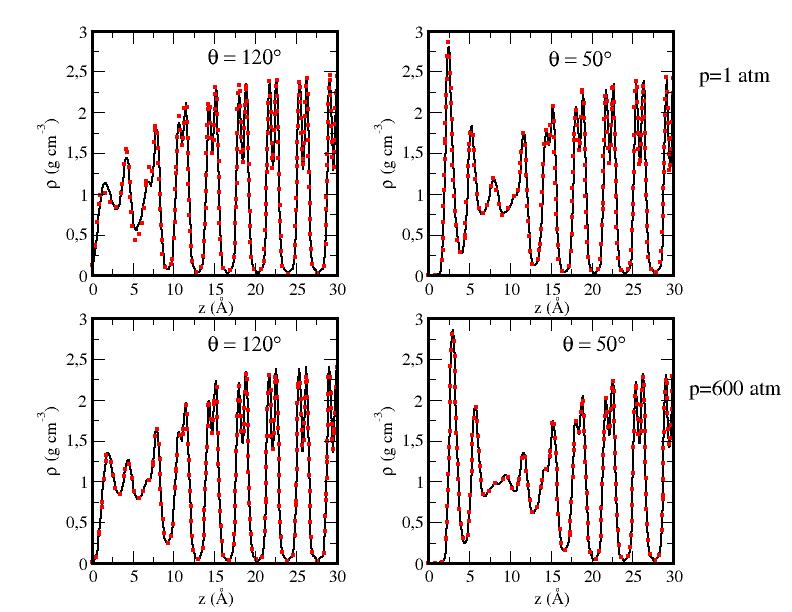}
	\caption{Influence of surface preparation on interfacial premelting. The full lines correspond to
	   equilibrium density profiles obtained from an ice slab with a half
	   terminated bilayer as the initial configuration. The symbols correspond
	   to equilibrium density profiles obtained from simulations with an
	   equilibrated ice/vapor interface as the initial configuration. Results
	   are shown for $T=262$~K, with pressures of 1 and 600~atm and contact
	   angles of $\theta=120^\circ$ and $50^\circ$ as indicated in the panels.
	}
	\label{comparison_total_densityp}
\end{figure}

\begin{figure}[htb!]
	\centering
	\includegraphics[width=0.48\textwidth]{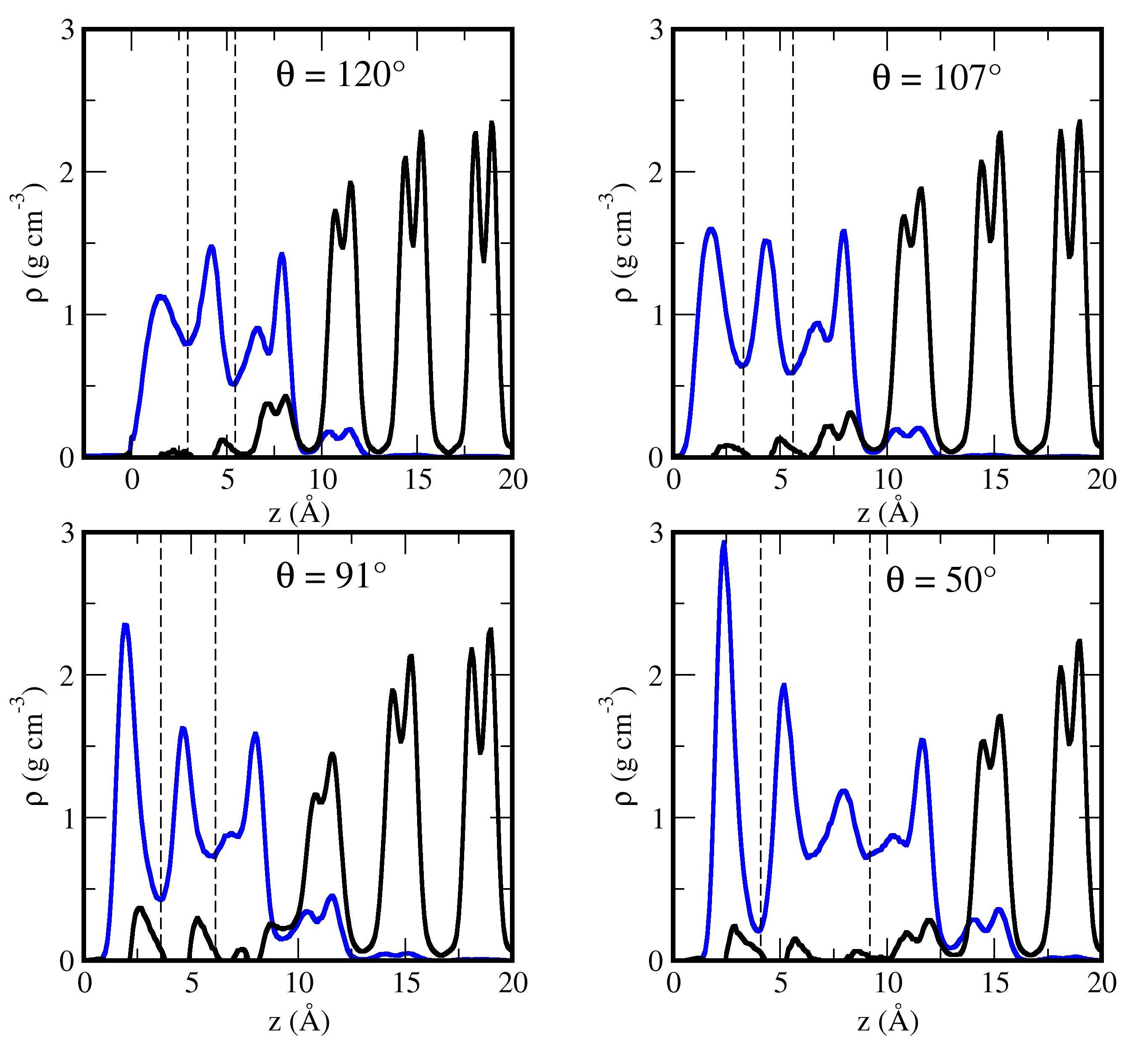}
	\includegraphics[width=0.48\textwidth]{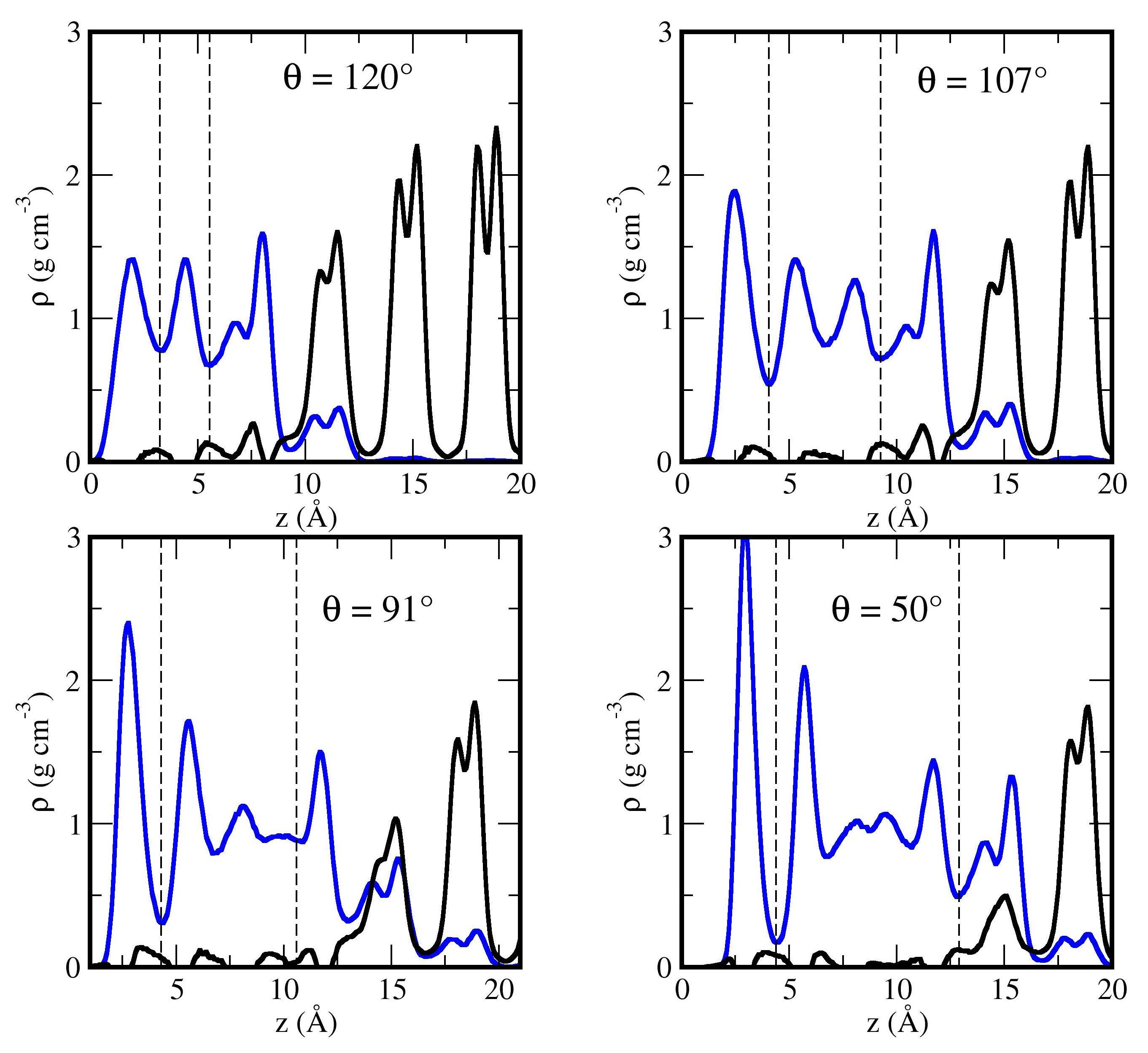}
	\caption{
		Structure of the interfacial premelting films at $p=1$~atm (left)
		and $p=600$~atm
		(right). The solid blue lines are the densities of liquid-like molecules, and
		the black lines correspond to densities of solid-like molecules. Notice a small
		penetration of solid-like molecules into the interfacial premelting film. 
		Films at $p=600$~atm are significantly thicker than those found at
		$p=1$~atm.
	} 
	\label{Fdensity_il}
\end{figure}

\begin{figure}[h!]
	\centering
	\includegraphics[width=0.7\textwidth]{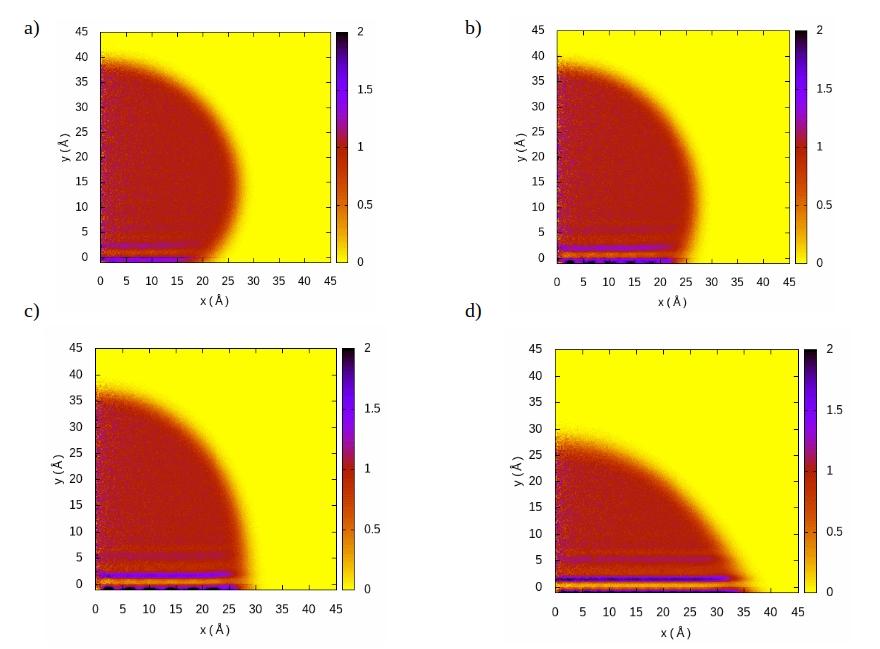}
	\caption{Density profiles $\rho(x,y)$ for sessile drops comprised of 2304 water molecules for interaction
		strength $f=1$ \textbf{(a)}, $f=2$ \textbf{(b)}, $f=3$ \textbf{(c)}, and $f=4$ \textbf{(d)}. Color bar depicts the density in the units of g cm$^{-3}$.}
\end{figure}

\newpage

\begin{figure}[p!]
	\begin{center}
		\includegraphics[width=0.7\textwidth]{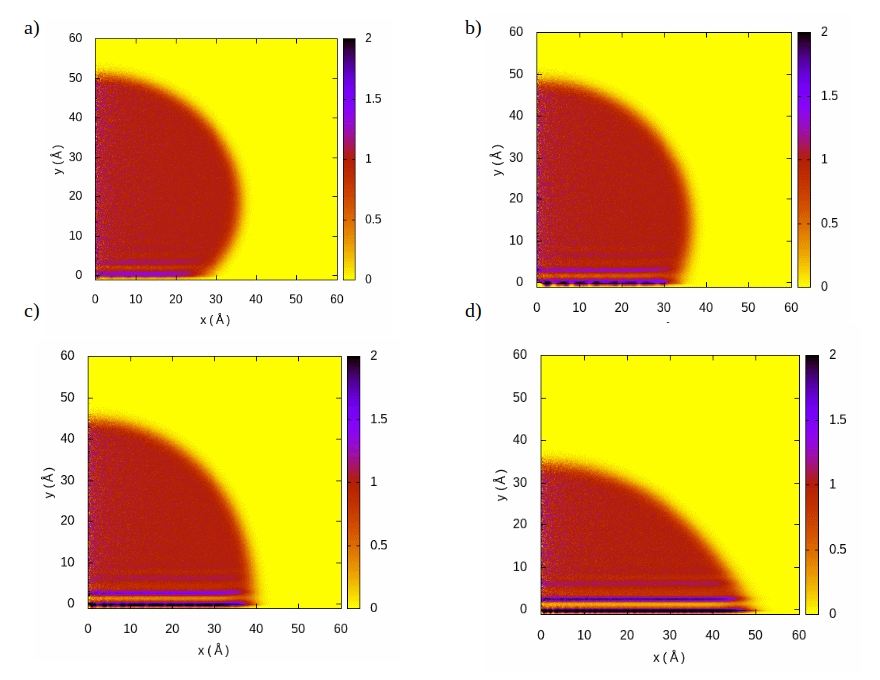}
		\caption{Density profiles $\rho(x,y)$ for sessile drops comprised of 5120 water molecules for interaction
			strength $f=1$ \textbf{(a)}, $f=2$ \textbf{(b)}, $f=3$ \textbf{(c)}, and $f=4$ \textbf{(d)}. Color bar depicts the density in the units of g cm$^{-3}$.}
			\end{center}
		\end{figure}
		
\begin{figure}[p!]
	\begin{center}		
		\includegraphics[width=0.6\textwidth]{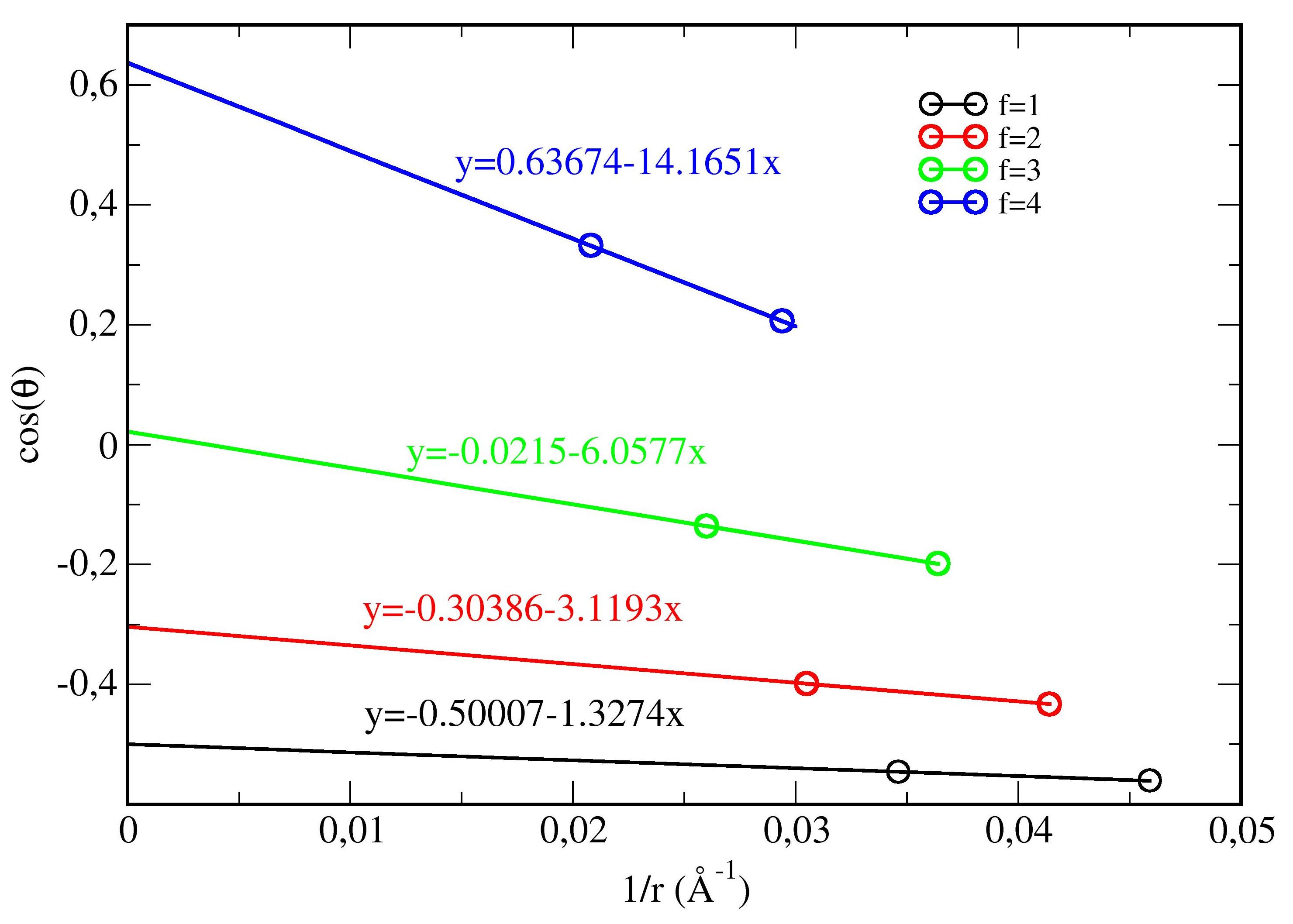}
		\caption{Measuring macroscopic contact angles. The finite size contact angles
			are extrapolated to infinite size in a plot of $\cos(\theta)$ vs
			$1/r$ where $r=R \sin(\theta)$.\label{cost}}
	\end{center}
\end{figure}


\clearpage
\newpage

\clearpage

\newpage

\begin{table}[htb!]
	\begin{center}
		\begin{tabular} {|cc|cc|ccccc|}
			\hline
			f  & p (atm) & $D$ ($10^{-10}$m$^2\cdot$s$^{-1})$ & $\eta$ (mPa$\cdot$s) & $z_w$ (\AA) & $\sigma_1$ (\AA) & $\sigma_2$ (\AA) & $u_s/U$ & $\tau$ (atm) \\
			\hline
			
			1  &  1 &  2.4 & 8.4 & -1.89 & 0.96 &   7.6  &  0.1   &  70   \\
			2  &  1 &  2.4 & 8.4 & -1.16 &  1.3 &   8.2  &  0.5   & 301   \\
			3  &  1 &  2.4 & 8.4 & -0.65 &  2.  &   10   &  0.85  & 485   \\
			4  &  1 &  2.4 & 8.4 & -0.16 &  2.6 &   12.5 &  1.    & 460   \\
			\hline
			1 &  600& 3.5 & 5.4 & -1.35 &  1.1  &  10.1  &   0.25 &   118   \\
			2 & 600 & 3.5 & 5.4 & -0.41 &  2.0  &  12.5  &   0.9  &   301   \\
			3 & 600 & 3.5 & 5.4 &  0.06 &  3.   &  15.3  &   0.95 &   305   \\
			4 & 600 & 3.5 & 5.4 &  0.31 &  4.5  &  15.4  &   1.   &   360   \\
			\hline
		\end{tabular}
	\end{center}
	\caption{Measure of hydrodynamic boundaries in the sliding experiments at $T=262$ K and various pressures. The simulation setup has the origin of coordinates close but not exactly at the wall position, which is allowed to move freely. $z_w$ is the average   wall position; $\sigma_1$ and $\sigma_2$ are the hydrodynamic boundaries   estimated from visual inspection of the velocity profiles and used to measure   the Couette thickness, $d_c=\sigma_2-\sigma_1$.  $u_s$ is the flow velocity   at $\sigma_1$ in units of the slider velocity, $U$. The table also includes   bulk diffusion coefficient and viscosity as obtained from equilibrium simulations.
	}
	\label{boundary}
\end{table}

\newpage

\begin{table}[htb!]
	\begin{center}
		\begin{tabular} {|cc|cc|ccccc|}
			\hline
			f  & T (K) & $D$ ($10^{-10}$m$^2\cdot$s$^{-1})$ & $\eta$ (mPa$\cdot$s) & $z_w$ (\AA) & $\sigma_1$ (\AA) & $\sigma_2$ (\AA) & $u_s/U$ & $\tau$ (atm)   \\
			\hline
			
	1  &  230 &  0.045 &  981   & -2.33 & -  & -   & -    &  36 \\
	4  &  230 &  0.045 &  981   & -1.2 &  1.95 &   9.36 &  1.  &  1181   \\
			\hline
	1 & 240 & 0.26 & 113 &  -2.52 &   -       &   -     &  - & 38   \\
	4 & 240 & 0.26 & 113 &  -0.92 &  2.14  & 9.33  &   1.   & 994   \\
			\hline
	1 & 250 & 0.95 & 24.0 & -2.35 &  -    &  -   &  - &    54 \\
	4 & 250 & 0.95 & 24.0 & -0.55 &  2.31  &  10.01  &   1.    & 778  \\
			\hline
      1 & 266 & 3.24 & 6.47 & -1.86 &  1.9  &  8.9 &   0.11 &    58   \\
	4 & 266 & 3.24 & 6.47 & -0.12 &  2.65  &  13.  &   1.   &   356  \\
			\hline
		\end{tabular}
	\end{center}
	\caption{Measure of hydrodynamic boundaries in the sliding experiments at various temperatures all at $p=1$ atm and various temperatures. The simulation setup has the origin of coordinates close but not exactly at the wall position, which is allowed to move freely. $z_w$ is the average   wall position; $\sigma_1$ and $\sigma_2$ are the hydrodynamic boundaries   estimated from visual inspection of the velocity profiles and used to measure   the Couette thickness, $d_c=\sigma_2-\sigma_1$.  $u_s$ is the flow velocity   at $\sigma_1$ in units of the slider velocity, $U$. The table also includes bulk diffusion coefficient and viscosity as obtained from equilibrium simulations.
	}
	\label{boundary_temps}
\end{table}

	\newpage

\begin{table}[htb!]
\begin{center}
\begin{tabular} {|c|c|c|c|c|c|}
 \hline
 p (atm) & f (kcal mol$^{-1}$) & $p_z\mathrm{equil}$ (atm) & $p_z\mathrm{shear, 5m/s}$ (atm) & $p_x\mathrm{shear, 5m/s}$ (atm) &  $p_x\mathrm{shear, 0.5m/s}$ (atm)\\
\hline
 \multirow{4}{*}{1} & 1.0 & $18.9$ & $3.4$ & $70.4$ & \\
 & 2.0 & $28.0$  & $4.7$ & $301$ &\\
 & 3.0 & $26.0$ & $9.6$ & $485$ &\\
 & 4.0 & $24.0$ & $7.2$ & $460$ &\\
\hline
\multirow{4}{*}{200} & 1.0 & $196$ & & &\\
 & 2.0 &  $208$& & &\\
 & 3.0 &  $214$& & &\\
 & 4.0 & $194$  & & &\\
\hline
\multirow{4}{*}{400} & 1.0 & $389$ & & &\\
 & 2.0 &  $410$& & &\\
 & 3.0 &  $410$& & &\\
 & 4.0 & $418$  & & &\\
\hline
\multirow{4}{*}{600} & 1.0 & $604$ & $ 601$ & $118$ & $22.2$ \\
 & 2.0 &  $592$& $605$ & $302$ & $39.7$\\
 & 3.0 &  $562$& $581$ & $305$ & $43.7$\\
 & 4.0 & $610$  & $595$ & $360$ & $43.0$ \\
\hline
\multirow{4}{*}{700} & 1.0 & $705$ & & &\\
 & 2.0 &  $681$& & &\\
 & 3.0 &  $671$& & &\\
 & 4.0 & $705$  & & & \\
\hline
\multirow{4}{*}{800} & 1.0 & $800$ & & &\\
 & 2.0 &  $797$& & &\\
 & 3.0 &  $808$& & &\\
 & 4.0 & $827$  & & &\\
\hline
\end{tabular}
\end{center}
\caption{Measure of stress components at the wall in equilibrium and shear
   simulations at $T=262$ K. The comparison of target pressure and perpendicular pressure
   measured at the wall serves to gauge the barostat. Results for the lateral
   force exterted by the wall on the water molecules allows to gauge the shear
   stress.
}
\label{stress}
\end{table}

\newpage

\begin{table}[htb!]
 \begin{center}
\begin{tabular} {|c|c|c|c|c|c|}
 \hline
  temperature (K) & f & $p_z\mathrm{equil}$ (atm) & $p_z\mathrm{shear, 5m/s}$ (atm) & $p_x\mathrm{shear, 5m/s}$ (atm) \\
 \hline
 \multirow{2}{*}{230} & 1.0 & 6.83 & 3.61 & 36.0 \\
                    & 4.0 & 24.0 & 13.9 & 1181\\
 \hline
 \multirow{2}{*}{240} & 1.0 & 5.93 & 4.35& 38.5 \\
                    & 4.0 & 10.7 & 11.3 & 994\\
 \hline
 \multirow{2}{*}{250} & 1.0 & 5.65 & 1.61 & 54.4 \\
                    & 4.0 & 17.1 & 8.70 & 778\\
 \hline
 \multirow{2}{*}{266} & 1.0 & 9.50 & 3.04 & 58.6 \\
                    & 4.0 & 4.76 & 12.3 & 356 \\
 \hline
\end{tabular}
\caption{Measure of stress components at the wall in equilibrium and shear
   simulations at $p=1$ atm. The comparison of target pressure and perpendicular pressure
   measured at the wall serves to gauge the barostat. Results for the lateral
   force exerted by the wall on the water molecules allows to gauge the shear
   stress.}
\label{stress_temps}
\end{center}
\end{table}

\newpage

\begin{table}[h!]
\begin{center}
\begin{tabular} {|c|c|c|c|c|c|c|}
 \hline
No. of water mols. & f (kcal mol$^{-1}$) & $\theta$ (deg) & $\cos(\theta)$ & R (\AA) & r (\AA) 
& 1/r (\AA$^{-1}$)\\
\hline
 \multirow{4}{*}{2304} 
 & 1.0 & 124.16  & -0.561 & 26.36 & 21.80 & 0.0459 \\
 & 2.0 & 115.64 & -0.433 & 26.8 & 24.15 & 0.0414\\
 & 3.0 & 101.47 & -0.199 & 28.04 & 27.48 & 0.0364\\
 & 4.0 & 78.13   & 0.206 & 34.76 & 34.02 & 0.0294\\
 \hline
  \multirow{4}{*}{5120} 
 & 1.0 & 123.12 & -0.546 & 34.56 & 28.94 & 0.0346 \\
 & 2.0 & 113.5 & -0.399 & 35.7 & 32.74 & 0.0305\\
 & 3.0 & 97.8  & -0.136 & 38.75 & 38.39 & 0.0260 \\
 & 4.0 & 70.6  &  0.332 & 50.9 & 48.01 & 0.0208 \\
 \hline
 \end{tabular}
\caption{The relation of contact angle values with respect
to the interaction strength $f$ for two system sizes simulated. }
\label{tab6}
\end{center}
\end{table}


\clearpage

\section*{Supplementary Movies}

\begin{figure}[htb!]	
	\includegraphics[width=0.15\textwidth]{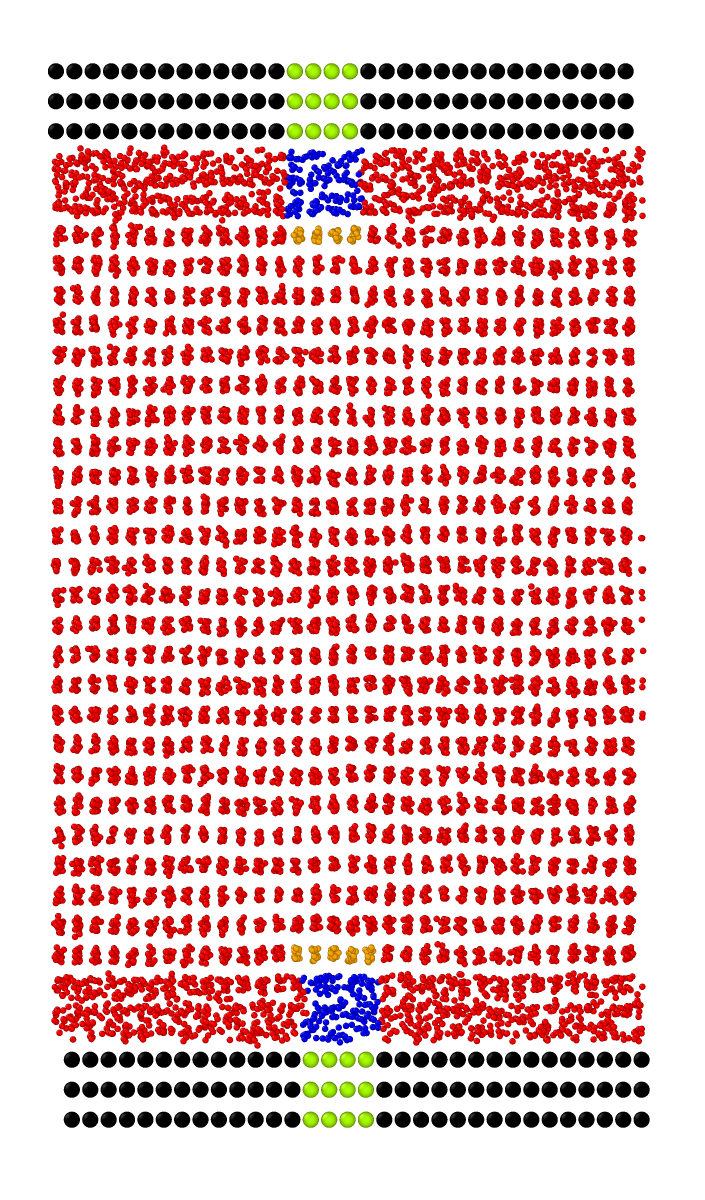}
	\caption*{{\bf Movie S1}: Sliding with slip at a hydrophobic wall. Movie
	displays the first 0.5~ns of a simulation at p=1~atm and T=262~K where a
   hydrophobic substrate with $\theta=120^\circ$ slides at a velocity of $U=5$~m/s. Notice how the wall slips past the premelting film, as illustrated by the blue tagged molecules.}
\end{figure}

\begin{figure}[htb!]	
	\includegraphics[width=0.15\textwidth]{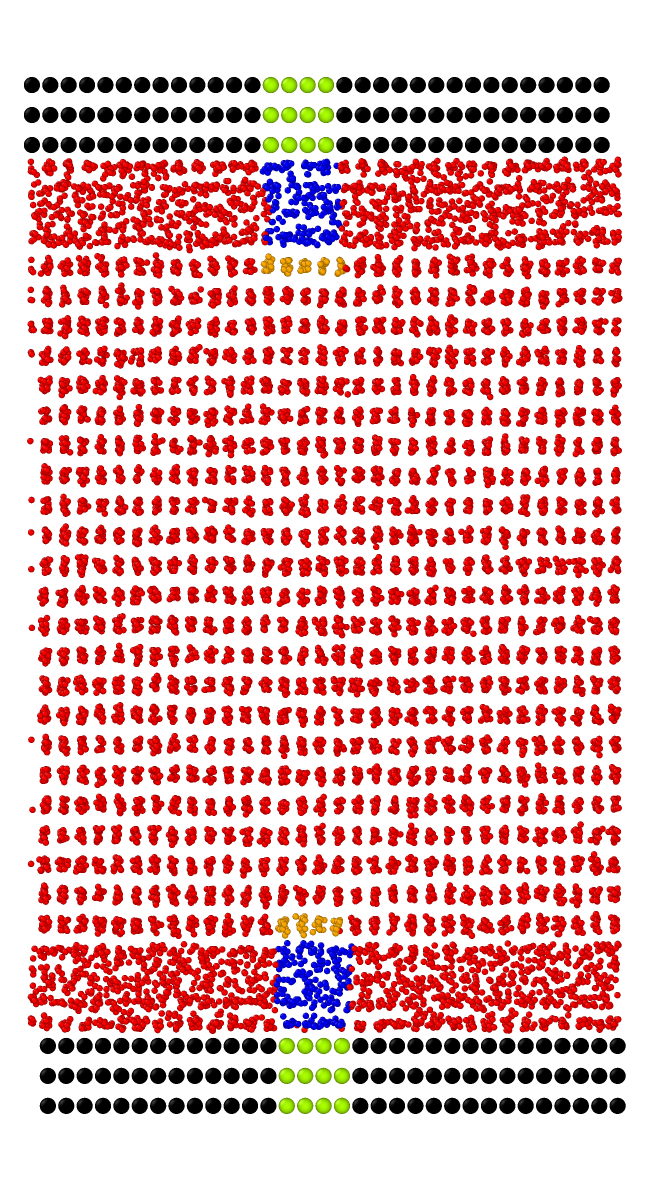}
	\caption*{{\bf Movie S2}: Sliding with stick at a hydrophobic wall. Movie
	   displays the first 0.5~ns of a simulation at p=1~atm and T=262~K where
	   a hydrophilic substrate with $\theta=50^\circ$ slides at velocity of
	   $U=5$~m/s. Notice how the the wall-adsorbed layer sticks to the wall and moves at equal speed, as indicated by the blue tagged moelcules.
	}
\end{figure}

\begin{figure}[htb!]	
	\includegraphics[width=0.15\textwidth]{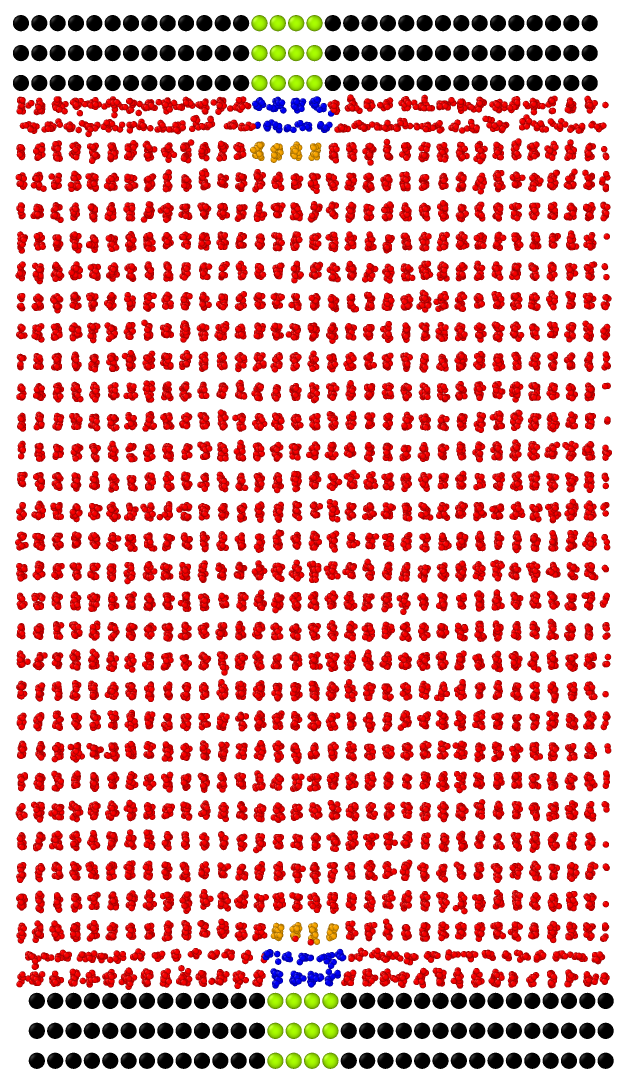}
	\caption*{{\bf Movie S3}: Frictional melting of ice. Movie displays
	   the full 10~ns of a simulation at p=1~atm and T=230~K where a
	   hydrophilic substrate with $\theta=50^\circ$ slides at velocity of
	   $U=5$~m/s. Notice how the first ice bilayer melts already at 0.5~ns and an additional bilayer has melted after 5~ns. 
	}
\end{figure}

\clearpage


\end{document}